\documentclass[3p]{elsarticle}

\usepackage{amssymb}
\usepackage{multirow}
\usepackage{multicol}
\usepackage{graphicx}
\usepackage{amsmath}
\usepackage{lineno}
\usepackage{setspace}
\usepackage{subcaption}
\usepackage{listings}
\usepackage{xspace}
\usepackage{xcolor}
\usepackage{float}

\newcommand{\pluseq}{\mathrel{+}=}
\newcommand{\plusplus}{\mathrel{+}+}
\newcommand{\tripoli}{{\text{Tripoli-4}}\textsuperscript{\textregistered}\xspace}
\newcommand{\mcnp}{\text{MCNP6}\xspace} 
\newcommand{\geant}{\text{Geant4}\xspace}
\newcommand{\njoy}{\text{NJOY}\xspace}

\journal{Annals of Nuclear Energy}

\begin{document}

\begin{frontmatter}

\title{Improvement of \geant Neutron-HP package: Doppler broadening of the neutron elastic scattering kernel and cross sections}

\author[1,2]{M. Zmeškal}\corref{cor1}
\ead{zmeskma1@fjfi.cvut.cz}
\author[3]{L. Thulliez}
\ead{loic.thulliez@cea.fr}
\author[3]{E. Dumonteil}

\address[1]{Dept. of Nuclear Reactors, Faculty of Nuclear Sciences and Physical Engineering, Czech Technical University in Prague, V Holesovickach 2, Prague, 180 00, Czech Republic}
\address[2]{Research Centre Rez, Hlavni 130, Husinec-Rez, 250 68, Czech Republic}
\address[3]{Université Paris-Saclay, CEA, Institut de Recherche sur les Lois Fondamentales de l'Univers, 91191, Gif-sur-Yvette, France}

\cortext[cor1]{Corresponding author}

\begin{abstract}

Whether it is for shielding applications or for safety criticality studies, numerically solving the neutron transport equation with a good accuracy requires to precisely estimate the Doppler broadened elastic scattering kernel in the thermal and epithermal energy range of neutrons travelling in a free gas. In \geant, low energy neutrons are transported using evaluated data libraries handled by the Neutron High-Precision (Neutron-HP) package. Version 11.00.p03 of the code features in particular the Doppler broadened elastic scattering kernel, provided by the so-called 'Sampling of the Velocity of the Target' (SVT) method. However this latter fails for resonant heavy nuclei such as $^{238}$U and can severely impact the solving of the Boltzmann equation in fissile media. To overcome this shortcoming, the Doppler Broadened Rejection Correction (DBRC) method has been implemented in \geant and successfully validated with the reference Monte Carlo neutron transport code \tripoli~(version~11). This development will be taken into account in the next release of the code. The cross section Doppler broadening process, which is performed on-the-fly, is also carefully investigated and ways to improve it on a simulation-by-simulation basis are presented. All the validations have been performed with an automated benchmark tool which has been designed to support the quality assurance of the \geant Neutron-HP package. This tool is currently available on an ad hoc Gitlab repository and will be included in \geant. 

\end{abstract}

\begin{keyword}

\geant \sep Neutron-HP \sep Neutron elastic scattering kernel \sep Doppler broadening \sep DBRC \sep \tripoli

\end{keyword}

\end{frontmatter}


\section{Introduction}
\label{sec:introduction}

Monte Carlo codes are routinely used for particle transport applications since they allow to track  particles individually and eventually to keep correlations between them. To this aim the \geant open-source toolkit has been developed by an international collaboration with the aim of supporting high-energy physics applications, as well as other domains, like space science, medical physics, engineering and nuclear  physics.\cite{Agostinelli2003,Allison2006,Allison2016}. In the past decades, the \geant physics range has been extended to lower energies, including low energy neutron transport (\textit{i.e.} below 20 MeV) through the development of the Neutron High-Precision (Neutron-HP) package which is mainly based on evaluated nuclear data libraries to accurately describe neutron-matter interactions. During the last years in particular, the Neutron-HP package has been graduously improved ~\cite{Mendoza2014,Mendoza2018,Tran2018a,Thulliez2022}, due to the increasing need for neutron transport capabilities in the broader context of multi-particle transport codes required by applications such as accelerator physics, medical imaging and radiotherapy, fundamental physics, nuclear industry, etc. These latest \geant developments place the code close to be on-par with reference neutron transport codes such as \tripoli \cite{Brun2015} or \mcnp \cite{Werner2018}. 
\newline 
\indent
Neutron transport simulations heavily rely on an accurate description of neutron-matter interactions and more specifically on a proper handling of the temperature of the material in which the neutron propagate because low energy neutrons (thermal and epithermal) are sensitive to molecular and atomic thermal motions. This latter cannot usually be neglected during neutron transport since it affects both the cross section and the final state of the reactions. Indeed the available energy in the centre of mass of the system depends on the relative velocity between the neutron and the moving target nucleus. This effect gives rise to the Doppler broadening process which can be written as:
\begin{equation}
    \sigma_T(v_n)=\frac{1}{v_n}\int v_r \sigma_{T=0 \text{K}}(v_r) \mathcal{M}(\mathbf{V}_t,T) d\mathbf{V}_t
    \label{eq:doppler}
\end{equation}
\noindent
with $\sigma_T$ the Doppler broadened cross section at temperature $T$, $\mathcal{M}(\mathbf{V}_t,T)$ the target velocity distribution, $v_r=\|\mathbf{v}_n-\mathbf{V}_t \|$ the relative velocity between the neutron, $\mathbf{v}_n$, and the target nucleus, $\mathbf{V}_t$.
The cross section Doppler broadening can be performed exactly for example with the \mbox{SIGMA-1} algorithm \cite{Cullen1979}. The definition of the final state of the reaction has to be coherent with the Doppler broadened cross section value. Often in Monte Carlo neutron transport codes the final state definition is divided in three domains depending on the neutron energy $E_n$, as sketched in Figure~\ref{fig:Geant4_modelRanges}. For~$E_n$$>$SVT\_E\_max, with SVT\_E\_max=400$k_B T$ ($\sim$10.5~eV at 300~K) and $k_B$ the Boltzmann constant, the target nuclei are considered at rest, leading to the use of the so-called 0~K asymptotic elastic scattering kernel. For $E_n$$<$SVT\_E\_max, the target nuclei are assumed to behave as a free gas, \textit{i.e.} moving freely relative to each other, and are described by a Maxwellian velocity distribution which is a valid approximation for many targets (in particular metallic), therefore in the following $\mathcal{M}(\mathbf{V}_t,T)$ is  set to be a Maxwell-Boltzmann distribution. The final state is then derived from Equation~\ref{eq:doppler} assuming a constant cross section over the relative velocity range leading to the so called 'Sampling of the Velocity of the Target' (SVT) method. Furthermore if the molecular and atomic bonds have a significant impact, \textit{i.e.} for $E_n$$<$S($\alpha$,$\beta$)\_E\_max with S($\alpha$,$\beta$)\_E\_max=4 eV in \geant, the free gas model does not hold true anymore and S($\alpha$,$\beta$) tables grasping the complex underlying physics of neutron-molecule or neutron-atom interactions have to be used to compute both the cross sections and scattering kernels. These tables are available for most materials used in nuclear reactor physics such as light water or uranium dioxide, since it can significantly impact the simulation predictions.
\newline \indent
The SVT assumption used to compute the final state for $E_n$$<$SVT\_E\_max is not valid for resonant heavy nuclei since their cross sections vary rapidly by a few orders of magnitude over a narrow energy range. To overcome this shortcoming, after a presentation in Section \ref{sec:benchmark_ans_tools} of the tools used in this work, in Section \ref{sec:DBRC} the Doppler Broadening Rejection Correction (DBRC) algorithm implementation and validation in \geant are presented. Then in Section \ref{sec:DB_cross section} the on-the-fly cross section Doppler broadening process in \geant is carefully studied and improvements are suggested.

\section{Benchmark methodology and tools}
\label{sec:benchmark_ans_tools}

The developments performed in this work have been done with \geant version~11.00.p03. The results are compared to the reference neutron transport codes \tripoli~(version 11) and \mcnp~(version~6.2). \tripoli is developed since the mid-1990s at CEA-Saclay (France) and is used as a reference code by the main French nuclear companies \cite{Brun2015}. \mcnp is developed at Los Alamos National Laboratory (United States) and is used worldwide as a reference neutron transport code for many applications involving neutrons. Both codes are used for criticality-safety system studies, depletion calculations and shielding applications. They benefit from a very large verification and validation database gathering more than 1000 experimental benchmarks and are regularly validated/qualified using inter-code comparisons \cite{intercomp}.
\newline
The evaluated data library ENDF/B-VII.1 \cite{Chadwick2011} is used in this work through the G4NDL4.5 library in \geant.
\newline
\indent
In the following, the developments performed in \geant are validated with a differential-microscopic benchmark referred to as the thin-cylinder benchmark \cite{Mendoza2014,Thulliez2022}. Its geometry is a thin cylinder with a radius $r$=1~$\mu$m and a length $L$=2~m. Mono-energetic neutrons, with a mean free path $\lambda_n$, are shot along the cylinder axis. The cylinder dimensions are chosen to have at least one collision ($\lambda_n<<L$) and to ensure that only one collision will occur in the cylinder ($r<<\lambda_n$). The neutron characteristics after the collision are tallied such as energy, angle, etc. It has to be mentioned that this thin-cylinder benchmark is a part of a broader set of automated benchmarks, which also includes an integral-macroscopic benchmark called the sphere benchmark in which the neutron flux is tallied within a 30 cm radius sphere having a mono-energetic and isotropic source at its center \cite{Thulliez2022}. These automated benchmarks have been developed to test and assess the quality of new \geant releases with respect to its Neutron-HP package. In this framework, once the \geant simulations are performed, this tool compares the \geant results to \tripoli or \mcnp results aggregated in a database and tests the statistical significance of the results to alert on potential problems. Finally it outputs a report summarising the results. This code is made available to the community on the following Gitlab repository \cite{MarekGitLab2023} and will be included to \geant.
\newline
\indent
 In this work, the thin-cylinder material is set to $^{238}$U because it has low energy resonances in its cross sections. The primary neutron energies are chosen to be below the first three resonances, \textit{i.e.} at $6.52$~eV, $20.2$~eV and $36.25$~eV. The temperature effects on the elastic scattering kernels have been investigated with the $^{238}$U material set at $300$~K, $600$~K and $1000$~K. For each temperature-energy combination $10^8$ neutrons are simulated.  
 
\section{Doppler broadened neutron elastic scattering kernel}
\label{sec:DBRC}

\subsection{Study of the SVT algorithm for heavy nuclei}

The effect of thermal motion on the final state definition is mainly important for elastic scattering, because it is a non-threshold and non-absorbing reaction. Therefore in many neutron transport codes the SVT algorithm is just applied to elastic scattering whereas in \geant it is applied to all reaction channels (elastic, capture, etc). In the following only the elastic scattering reaction is considered. 
\newline \indent
The target nucleus thermal motion changes the outgoing neutron energy and scattering angle probabilities. To take it into account, the sampling of the velocity of the target nucleus $\mathbf{V}_t$ has to be done in a way that preserves the reaction rate given by Equation \ref{eq:doppler} \cite{Coveyou1956}. This sampling can be transformed into a selection of the pair $(\mu,V_t)$, where $\mu$ is the cosine of the angle between the neutron and target nucleus velocities. By introducing the exact form of $\mathcal{M}(\mathbf{V}_t,T)$ into Equation~\ref{eq:doppler} and integrating it over the azimuthal angle, the following joint probability distribution can be written:

\begin{equation}
    p(V_t,\mu) dV_t d\mu=\frac{v_r\sigma_0(v_r) \frac{\beta^3}{\pi^\frac{3}{2}} V_t^2 e^{-\beta^2V_t^2} 2\pi }{v_n\sigma_T(v_n)}dV_t d\mu
    \label{eq:probability}
\end{equation}
with $\beta=\sqrt{\frac{M}{2k_{B}T}}$ and $M$ the target mass. In this distribution $\mu$ and $V_t$ are correlated and cannot be sampled independently. When the assumption is made that $\frac{\sigma_0(v_r)}{\sigma_T(v_n)}$ is constant over the relative velocity range $v_r$, the SVT joint probability distribution is obtained:

\begin{equation}
    p_\text{SVT}(V_t,\mu) dV_t d\mu = C \underbrace{\frac{d\mu}{2}}_{(A)} \underbrace{(v_n+V_t)\frac{4\beta^3}{\sqrt{\pi}}V_t^2 e^{-\beta^2 V_t^2}dV_t }_{(B)} \underbrace{ \left(\frac{v_r}{v_n+V_t} \right)}_{(C)}
    \label{eq:SVT}
\end{equation}
In the SVT algorithm, $\mu$ is firstly sampled uniformly between $[-1,1]$.
Then in sampling $V_t$ from (B), $V_t$ is sampled from $\frac{4\beta^3}{\sqrt{\pi}}V_t^2e^{-\beta^2V_t^2}dV_t$ with probability $P_1=\frac{v_n}{v_n+\frac{2}{\sqrt{\pi}\beta}}$ and then from $2\beta^4 V_t^3e^{-\beta^2V_t^2}dV_t$ with probability $P_2=1-P_1$ \cite{Coveyou1956}. Finally the pair $(\mu,V_t)$ is accepted with the probability given by (C) where $v_r=\sqrt{v_n^{2}+V_{t}^{2}-2v_nV_t\mu}$. 
\newline 
\indent
The SVT method in \geant has been recently revised and successfully benchmarked against \tripoli ~\cite{Thulliez2022}. In this work it has been again benchmarked with an emphasis on resonant heavy nuclei, here on $^{238}$U, to validate the SVT algorithm which is the building block of the DBRC method. The default value of SVT\_E\_max has been increased from $400 k_\text{B}T$ ($\sim$10.5 eV at 300 K) to 50~eV to use the SVT algorithm for the first three $^{238}$U resonances. The Figures - left column - \ref{fig:comparison1}, \ref{fig:comparison2} and \ref{fig:comparison3} show that the energy of secondary neutrons calculated with \geant agree well with \tripoli within the statistical uncertainties. There is also an overall good agreement with \mcnp for the energy-temperature combination for which the SVT algorithm is used in \mcnp. In fact the default energy threshold SVT\_E\_max=400$k_BT$ in \mcnp could not have been changed in this work. 
Theses comparisons ensure that the SVT method is well implemented in \geant and allows to validate once again the building block of the DBRC algorithm for resonant heavy nuclei that is now studied.

\subsection{DBRC method for resonant heavy nuclei}

\begin{figure}[htbp]
    \centering
    \includegraphics[width=0.9\textwidth]{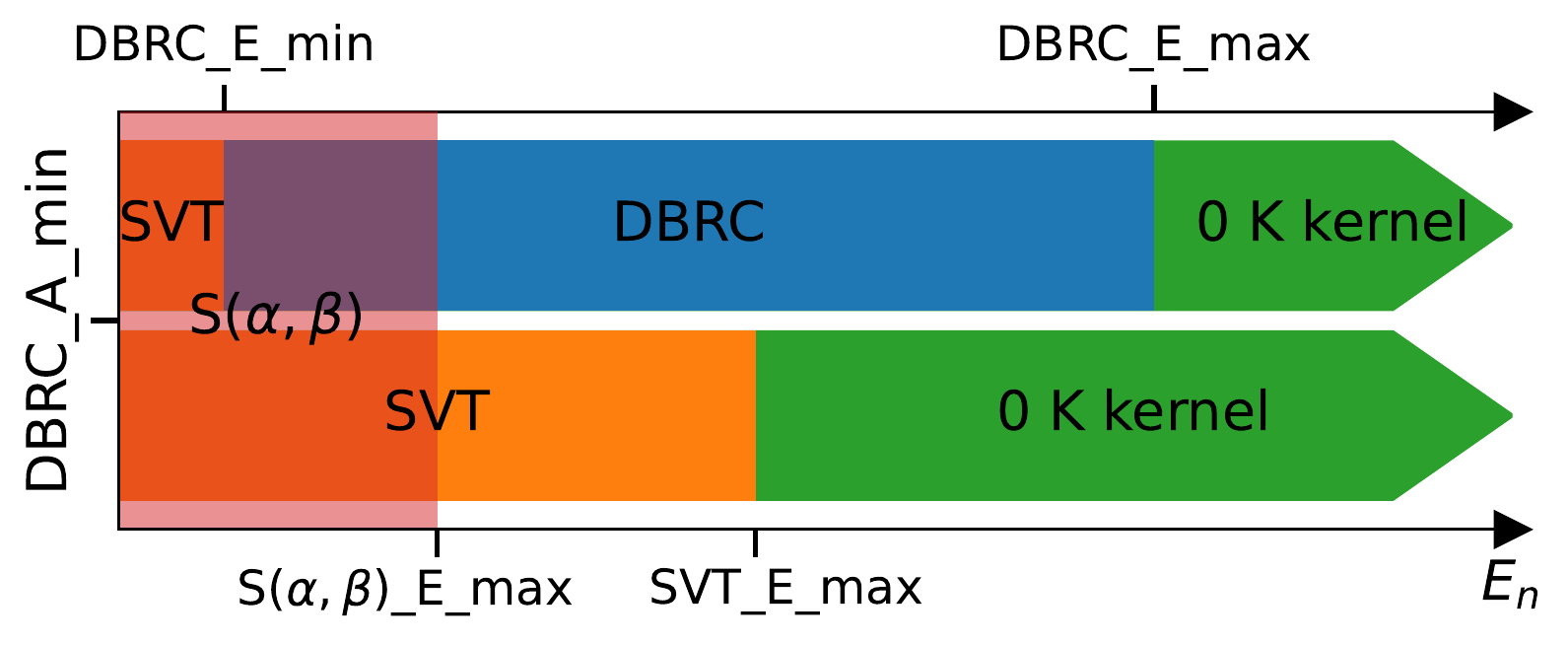}
    \caption{Doppler broadened elastic scattering final state models as a function of the neutron energy $E_n$. Bottom arrow : DBRC option switched off. Top arrow: DBRW option switched on. \newline    \textbf{SVT} stands for Sampling of the Velocity of the Target nucleus, \textbf{DBRC} for Doppler Broadening Rejection Correction, \textbf{\mbox{S($\alpha$, $\beta$)}} for tables grasping the complex underlying physics of neutron-molecule or neutron-atom interactions and \textbf{0~K kernel} for elastic scattering with the assumption of zero velocity of the target nucleus. The default threshold values in \geant are: \mbox{S($\alpha$, $\beta$)\_E\_max=4~eV}, SVT\_E\_max=400$k_BT$, DBRC\_E\_min=0.1~eV and DBRC\_E\_max=210~eV and DBRC\_A\_min=200~amu.} 
    \label{fig:Geant4_modelRanges}
\end{figure}

The assumption of the constant cross section leading to the SVT method is valid for light and medium nuclei since often their cross sections are constant at low energies. In the past it was assumed to be valid because it was thought that heavy nuclei do not significantly take part in the neutron slowing-down process. However it was shown \cite{Becker2009, Dagan2006, Dagan2011} that the effect of low energy resonances should not be overlooked in resonant heavy nuclei especially for shielding and nuclear reactor applications. In fact, for example, \mbox{Becker et al. \cite{Becker2009}} have demonstrated that treating exactly the elastic scattering kernel for the $^{238}$U resonances with DBRC instead of using the SVT assumption leads to a change of the k$_\text{eff}$ value by as much as $350$ pcm and Doppler coefficients by $16$ \% for an infinite array of identical fuel pin cells based on a PWR subassembly. It can also impact the extraction of resonance parameters from experimental data such as in \cite{Gunsing2012}.
\newline 
\indent
To overcome the SVT approximation shortcoming, in the mid-1990s, Rothenstein and Dagan \cite{Rothenstein1998,Rothenstein2004} successfully developed scattering kernels for resonant nuclei in the epithermal region through the extension of S($\alpha$, $\beta$) tables to higher energy. The DBRC algorithm that is now considered has been firstly proposed by Rothenstein \cite{Rothenstein1994a} and put again into light recently by Becker et al. \cite{Becker2009}. To exactly sample the ($\mu$, $V_t$) pair from Equation~\ref{eq:probability} on top of the SVT algorithm (Equation \ref{eq:SVT}) another rejection loop/criterion (D) is added to get the exact Doppler Broadened elastic scattering kernel~\cite{Rothenstein1995}:
\begin{equation}
    p_\text{DBRC}(V_t,\mu)=C' p_\text{SVT}(V_t,\mu)  \underbrace{\left( \frac{\sigma_{0}(v_r)}{\sigma_0^{\text{max}}(v_\xi)}\right)}_{(D)} 
    \label{eq:dbrc}
\end{equation}
with $C'$ a normalisation constant and $\sigma_{0}^\text{max}(v_\xi)$ the maximum $0$ K cross section in the range $v_n\pm 4\sqrt{\frac{2k_\text{B}T}{M}}$. This range is chosen to be in accordance with the integral limits of the \mbox{SIGMA-1} algorithm used for an exact Doppler broadening process \cite{Cullen1979}. 
From the rejection term (D), it can be seen, that the acceptance probability can be small near a resonance. Therefore this additional term increases the probability of selecting a velocity $V_t$ that makes the neutron scattered at the resonance energy. This leads to a computation time increase because of the low acceptance probability.
\newline \indent
The DBRC method has been implemented in G4ParticleHPElasticFS::GetBiasedThermalNucleus on top of the existing SVT algorithm as in previous works \cite{Becker2009,Sunny2012,Zoia2013,Brown2019}. To use and modify the behaviour of the DBRC algorithm in \geant, the users have access to five new commands allowing to modify its behaviour as sketched in Figure~\ref{fig:Geant4_modelRanges}:

\begin{itemize}
    \item /process/had/particle\_hp/use\_DBRC 
    \newline
    to enable (true) / disable (false) the DBRC algorithm. By default it is disabled.

    \item /process/had/particle\_hp/DBRC\_A\_min 
    \newline
    to set the minimum atomic mass in neutron masses for which the DBRC algorithm is applied. The default value is set to $200$~amu.
    
    \item /process/had/particle\_hp/SVT\_E\_max 
    \newline
    to set the maximum energy up to which the SVT algorithm is used. The default value is set to $400 k_\text{B}T$. The 0~K scattering kernel is used above the user defined value.  
    
    \item /process/had/particle\_hp/DBRC\_E\_min(max)  
    \newline
    to set the minimum (maximum) energy between which the DBRC algorithm is applied. The default value of E\_DBRC\_min is set to $0.1$~eV, below the SVT algorithm is used. This value should corresponds to an energy below which there is no nuclear resonance for the considered isotopes. The default value of E\_DBRC\_max is set to $210$~eV. These values are set in accordance with \cite{Sunny2012,Zoia2013}. 
\end{itemize}

The DBRC method being not implemented in the current \mcnp release, the \geant DBRC results are only compared to \tripoli. This latter has been validated by Zoia~et~al.~\cite{Zoia2013} who have performed an exhaustive validation of its implementation against other ways to compute the Doppler broadening elastic scattering kernel such as the use of S($\alpha$,$\beta$) tables extended to higher energies~\cite{Dagan2005}, the Weight Correction Method (WCM)~\cite{Lee2009}, or direct computation of the integral given by~\cite{Ouisloumen1991} in~\cite{Ghrayeb2011}.

\begin{figure}[htbp]
     \centering
     \begin{subfigure}[b]{0.49\textwidth}
         \centering
         \includegraphics[width=\textwidth]{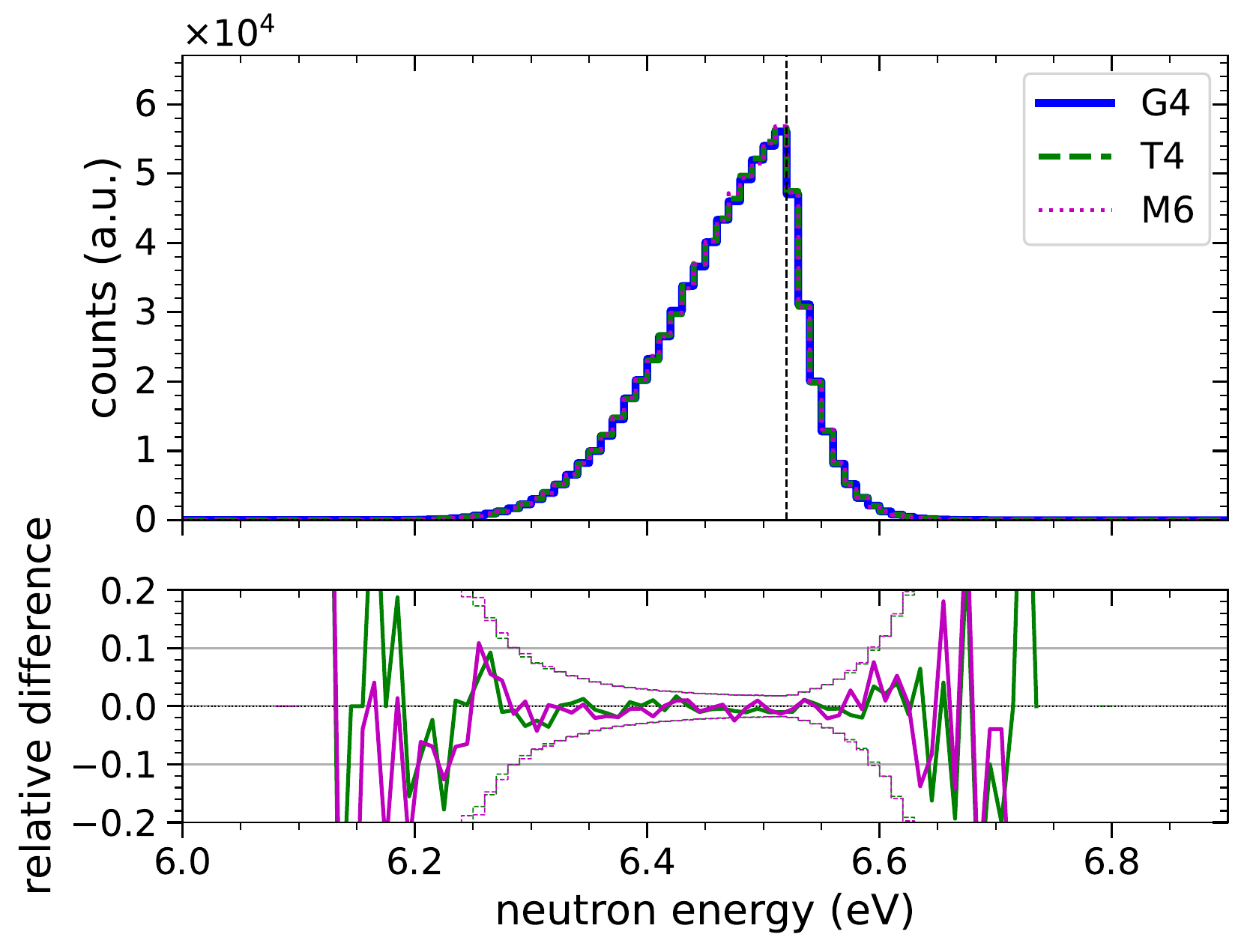}
         \caption{$6.52$ eV - $300$ K - SVT}
     \end{subfigure}
     \hfill
     \begin{subfigure}[b]{0.49\textwidth}
         \centering
         \includegraphics[width=\textwidth]{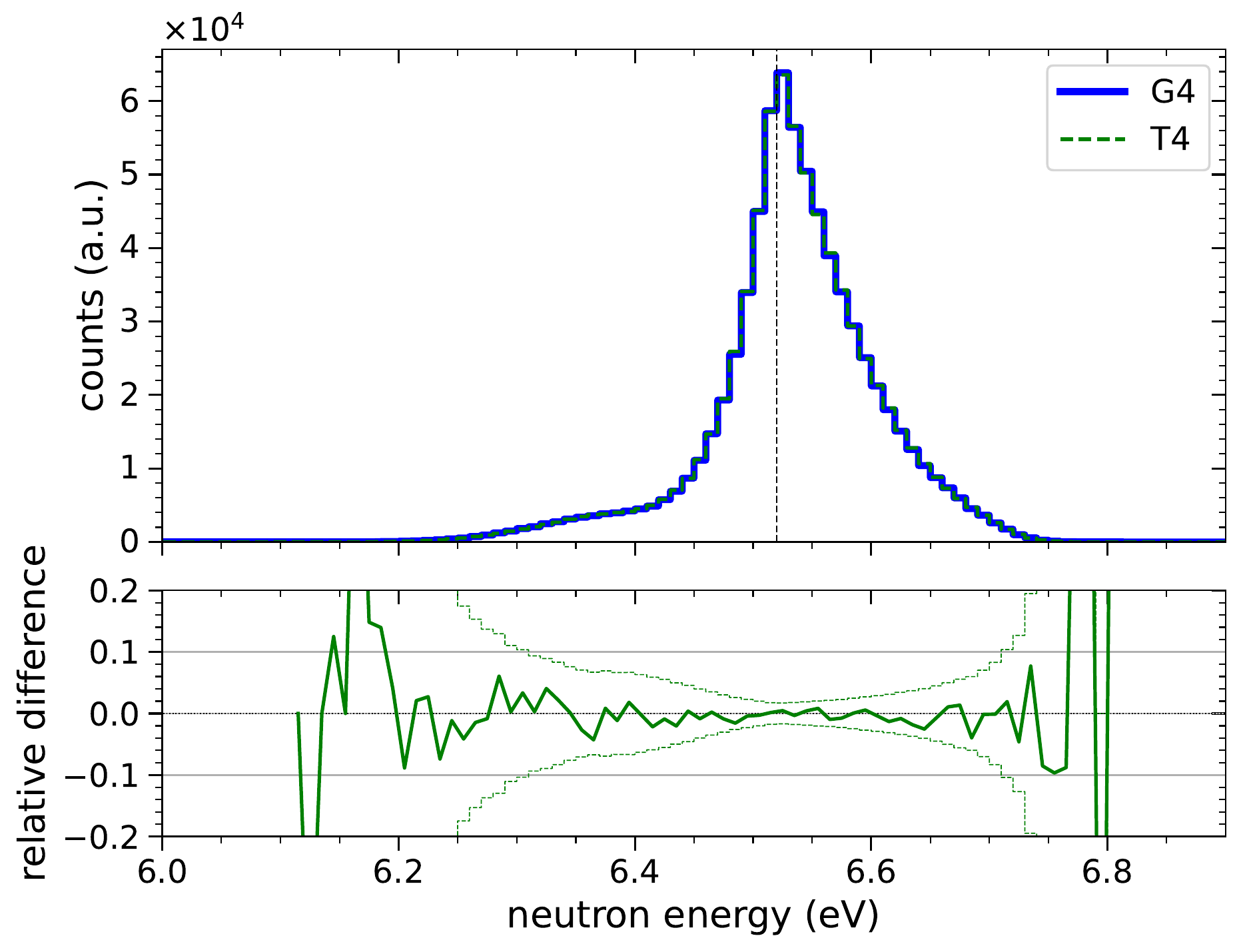}
         \caption{$6.52$ eV - $300$ K - DBRC}
     \end{subfigure}
     \begin{subfigure}[b]{0.49\textwidth}
         \centering
         \includegraphics[width=\textwidth]{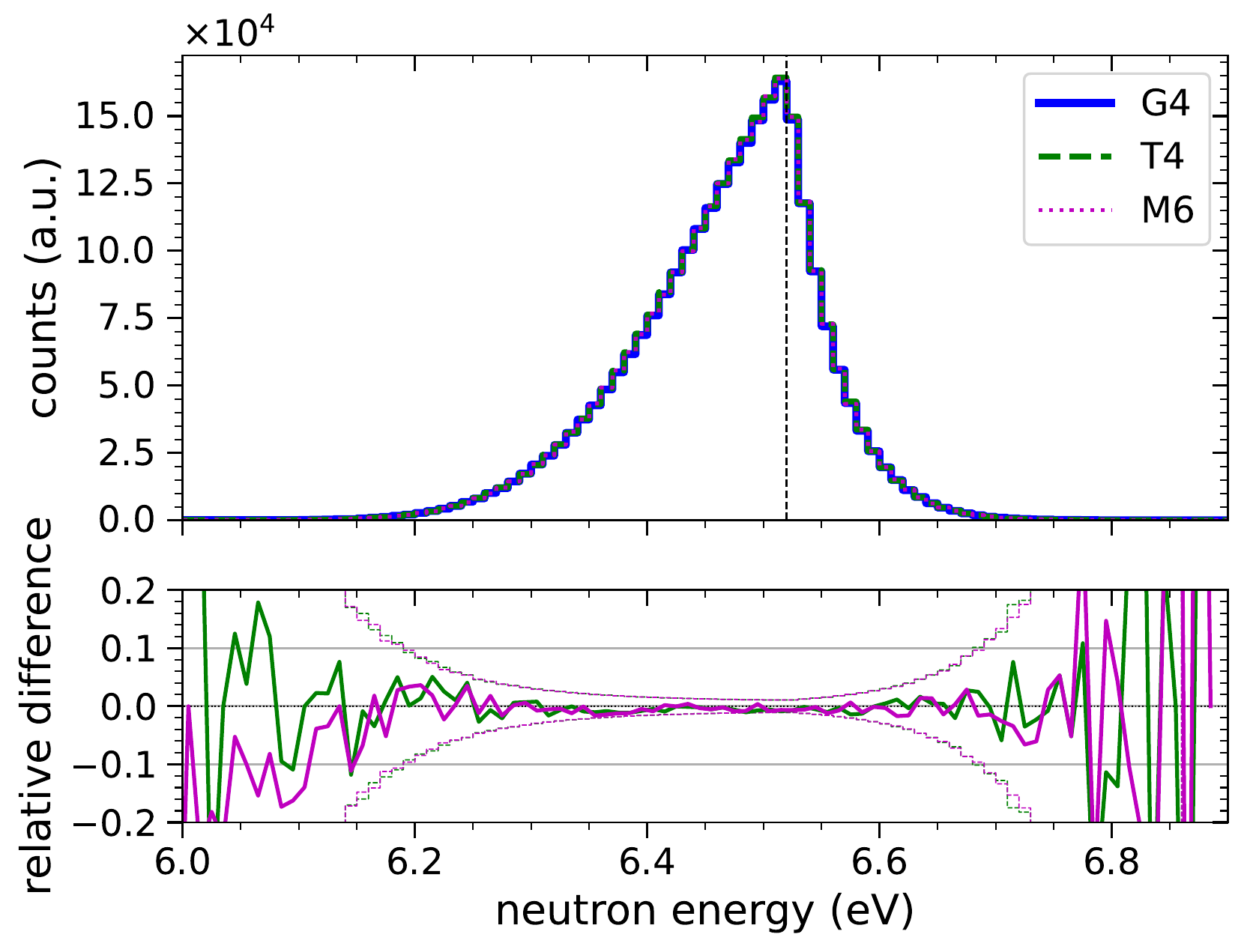}
         \caption{$6.52$ eV - $600$ K - SVT}
     \end{subfigure}
      \hfill
     \begin{subfigure}[b]{0.49\textwidth}
         \centering
         \includegraphics[width=\textwidth]{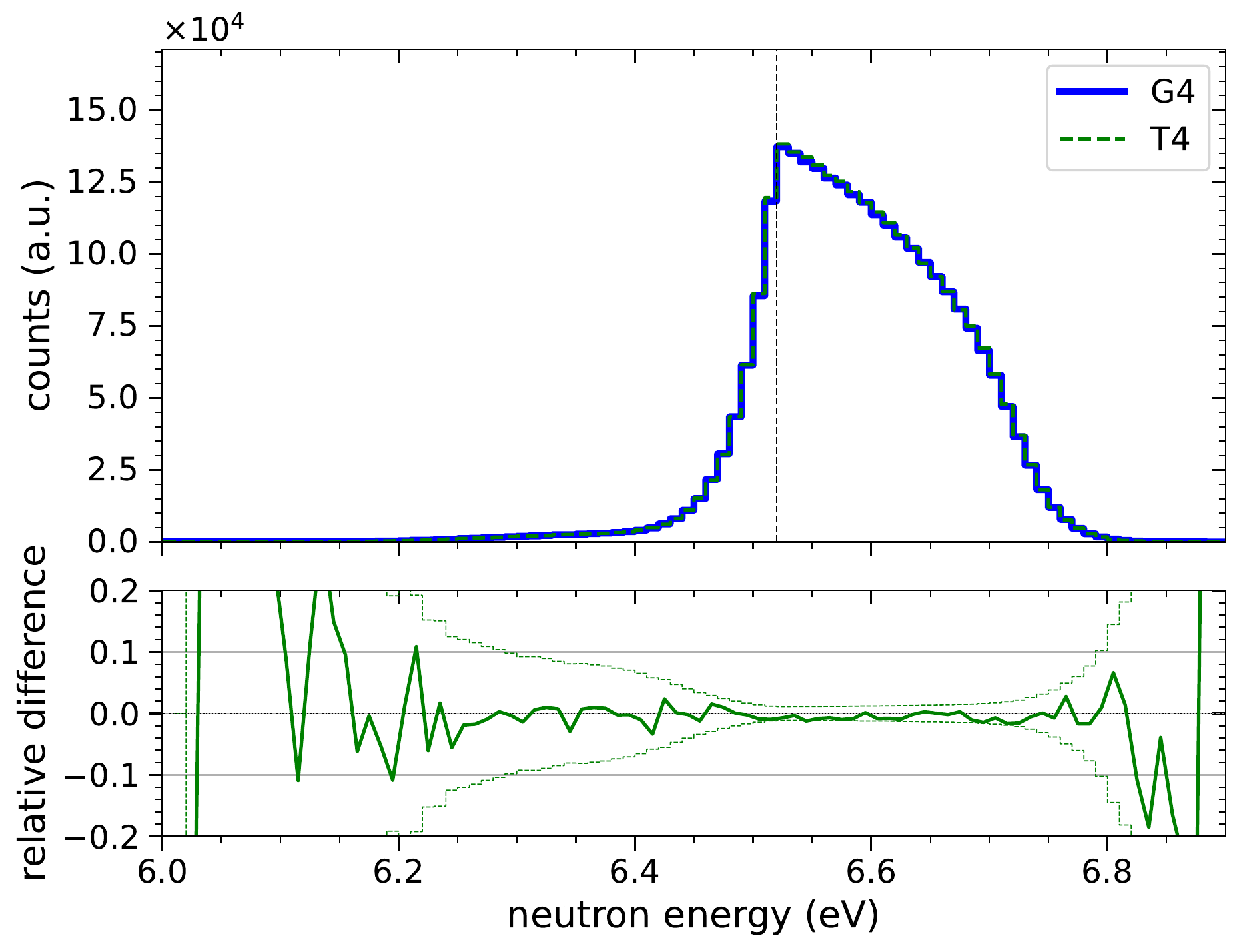}
         \caption{$6.52$ eV - $600$ K - DBRC}
     \end{subfigure}
     \begin{subfigure}[b]{0.49\textwidth}
         \centering
         \includegraphics[width=\textwidth]{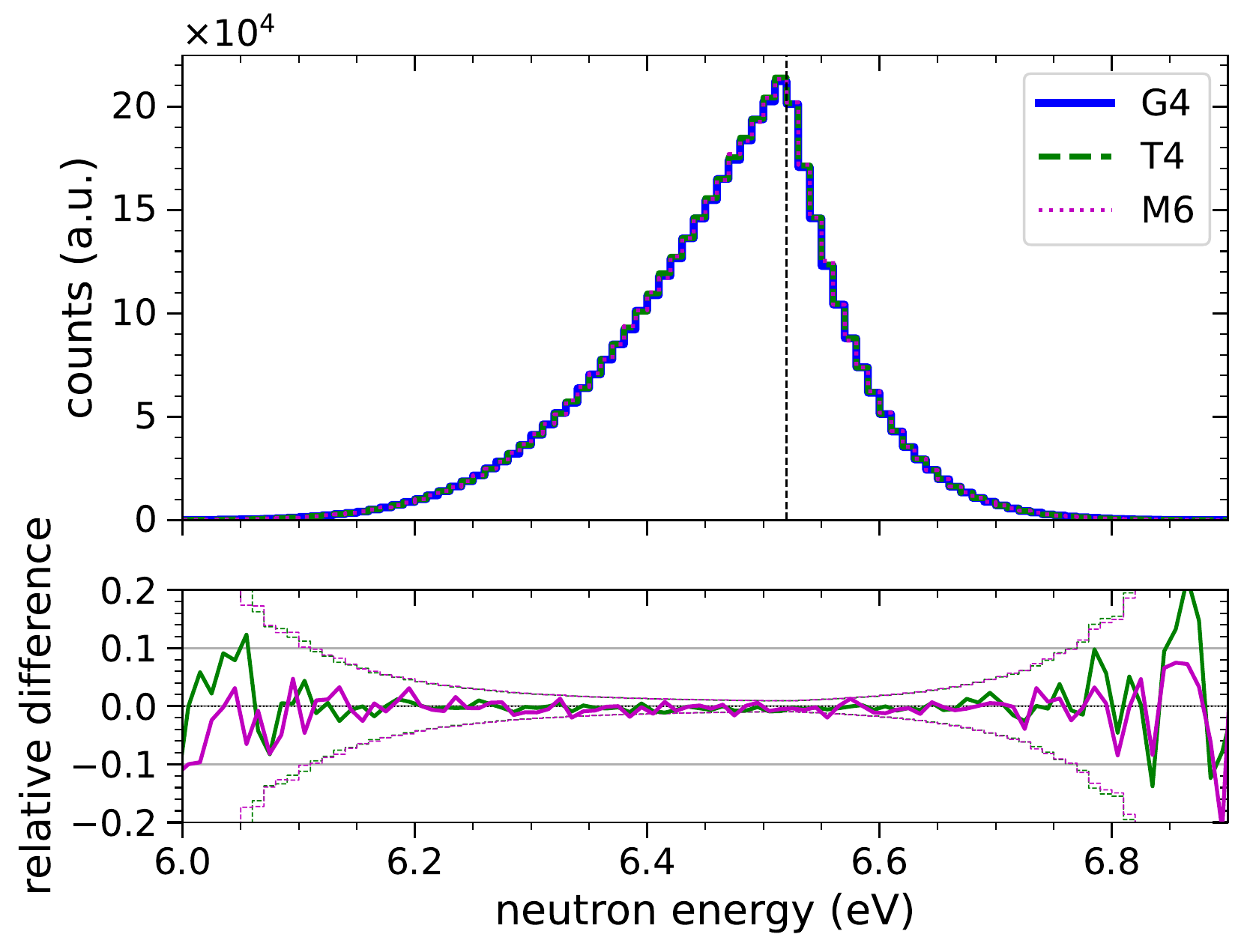}
         \caption{$6.52$ eV - $1000$ K - SVT}
     \end{subfigure}
     \hfill
     \begin{subfigure}[b]{0.49\textwidth}
         \centering
         \includegraphics[width=\textwidth]{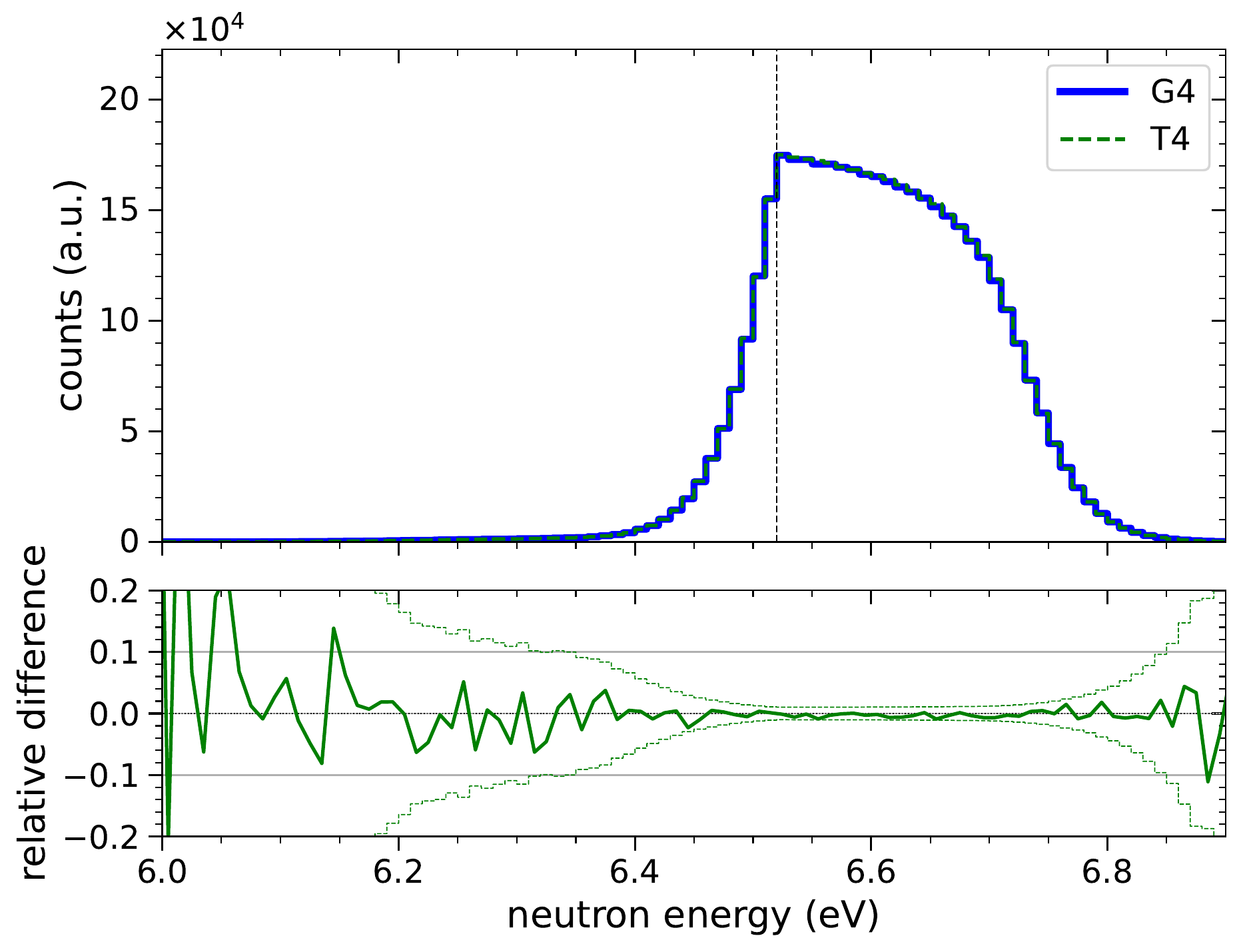}
         \caption{$6.52$ eV - $1000$ K - DBRC}
     \end{subfigure}
        \caption{Outgoing neutron energy distributions of scattered neutrons for $6.52$ eV neutrons on $^{238}$U at different temperatures computed with SVT (left column) and DBRC (right column) computed with \geant and \tripoli. The bottom part presents the relative difference G4/T4-1 (green solid line) and G4/M6-1 (magenta solid line) with three times the standard statistical error of this difference (dashed lines). \mcnp results at 300 K and 1000 K have been normalised to \tripoli ones, because the Doppler broadened cross sections at 300 K and 1000 K are effectively taken at 293.6 K and 900 K (pre-processed cross sections).}
        \label{fig:comparison1}
\end{figure}

\begin{figure}[htbp]
     \centering
     \begin{subfigure}[b]{0.49\textwidth}
         \centering
         \includegraphics[width=\textwidth]{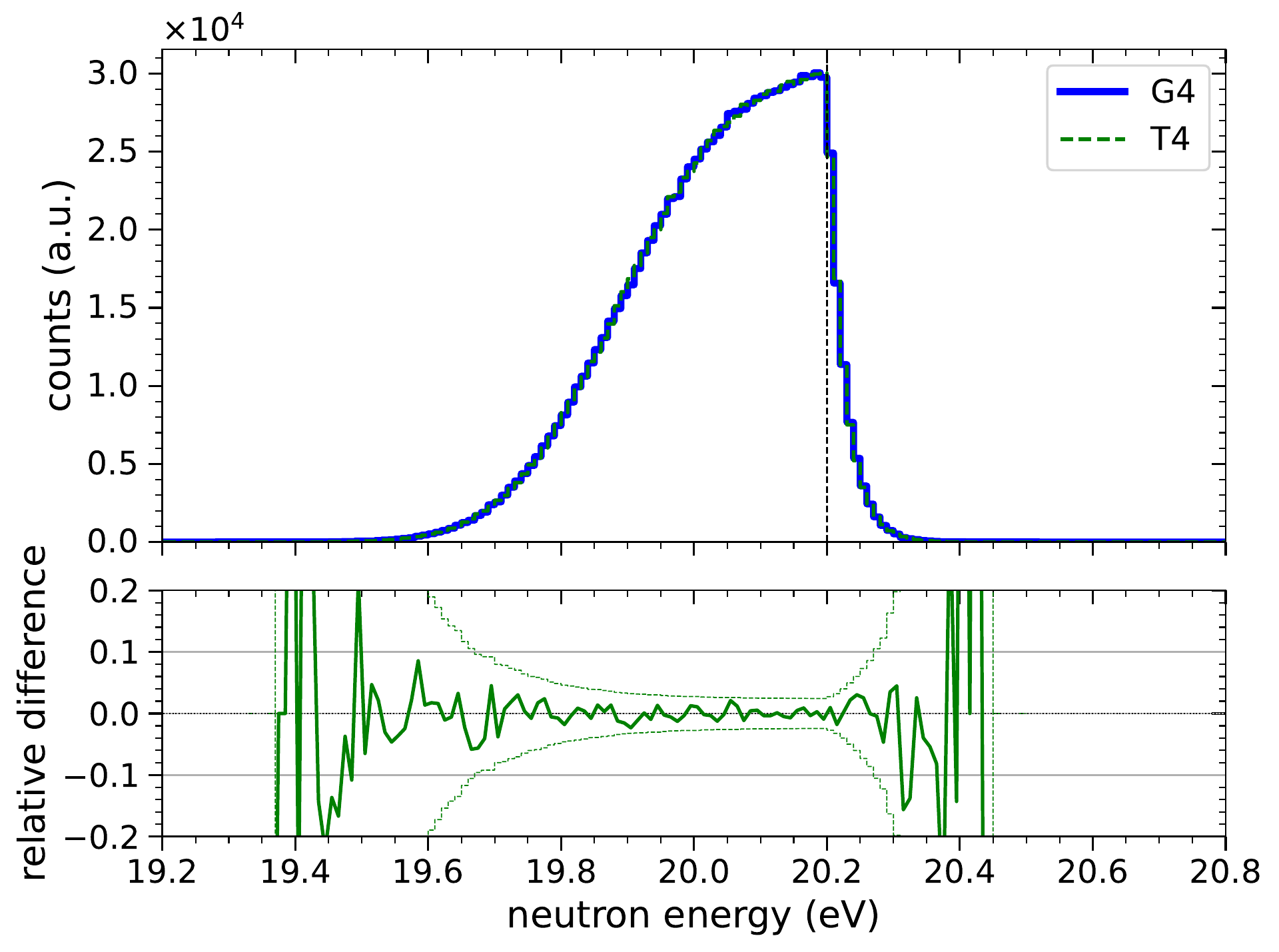}
         \caption{$20.2$ eV - $300$ K - SVT}
     \end{subfigure}
     \hfill
     \begin{subfigure}[b]{0.49\textwidth}
         \centering
         \includegraphics[width=\textwidth]{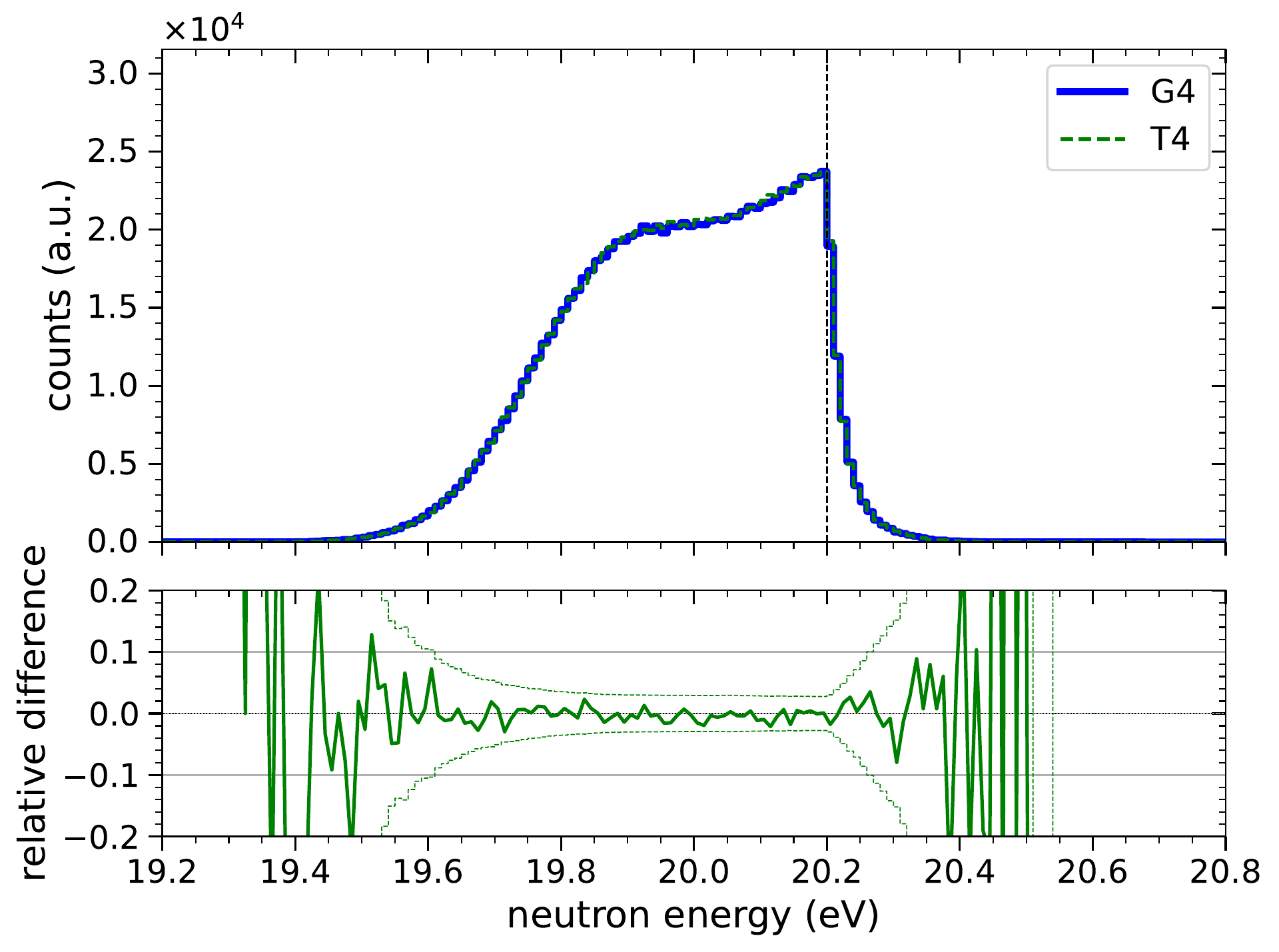}
         \caption{$20.2$ eV - $300$ K - DBRC}
     \end{subfigure}
     \begin{subfigure}[b]{0.49\textwidth}
         \centering
         \includegraphics[width=\textwidth]{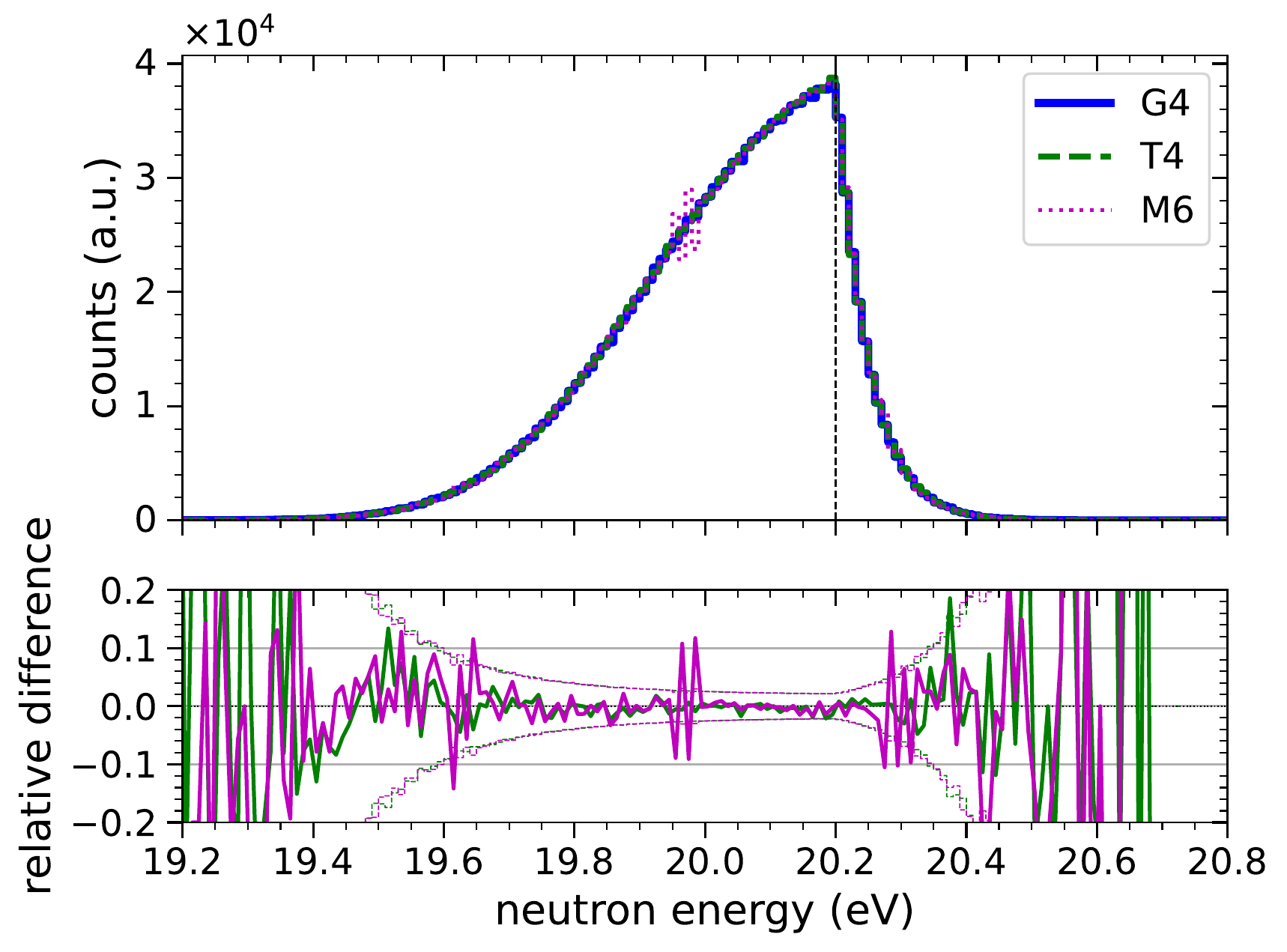}
         \caption{$20.2$ eV - $600$ K - SVT}
     \end{subfigure}
      \hfill
     \begin{subfigure}[b]{0.49\textwidth}
         \centering
         \includegraphics[width=\textwidth]{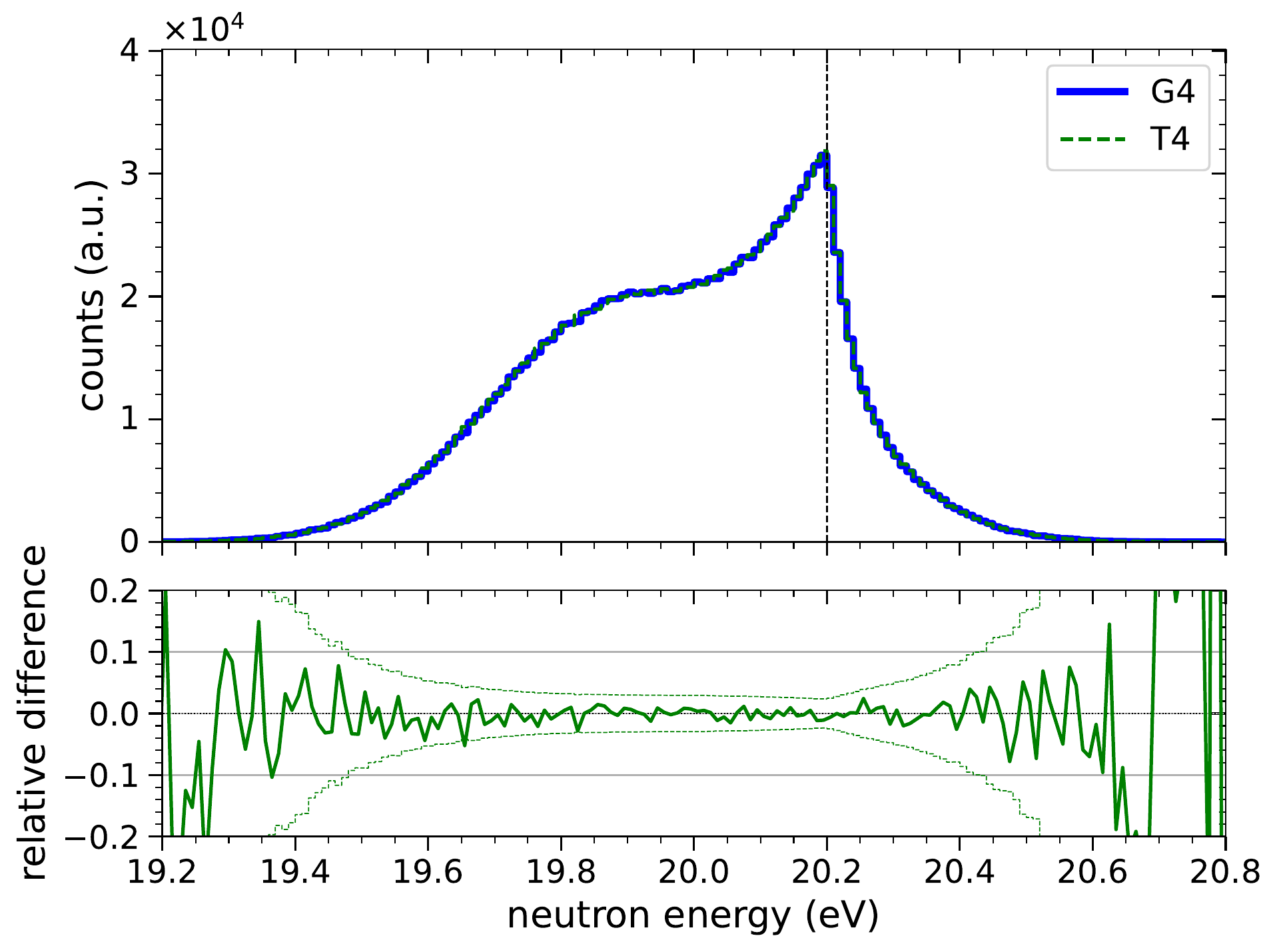}
         \caption{$20.2$ eV - $600$ K - DBRC}
     \end{subfigure}
     \begin{subfigure}[b]{0.49\textwidth}
         \centering
         \includegraphics[width=\textwidth]{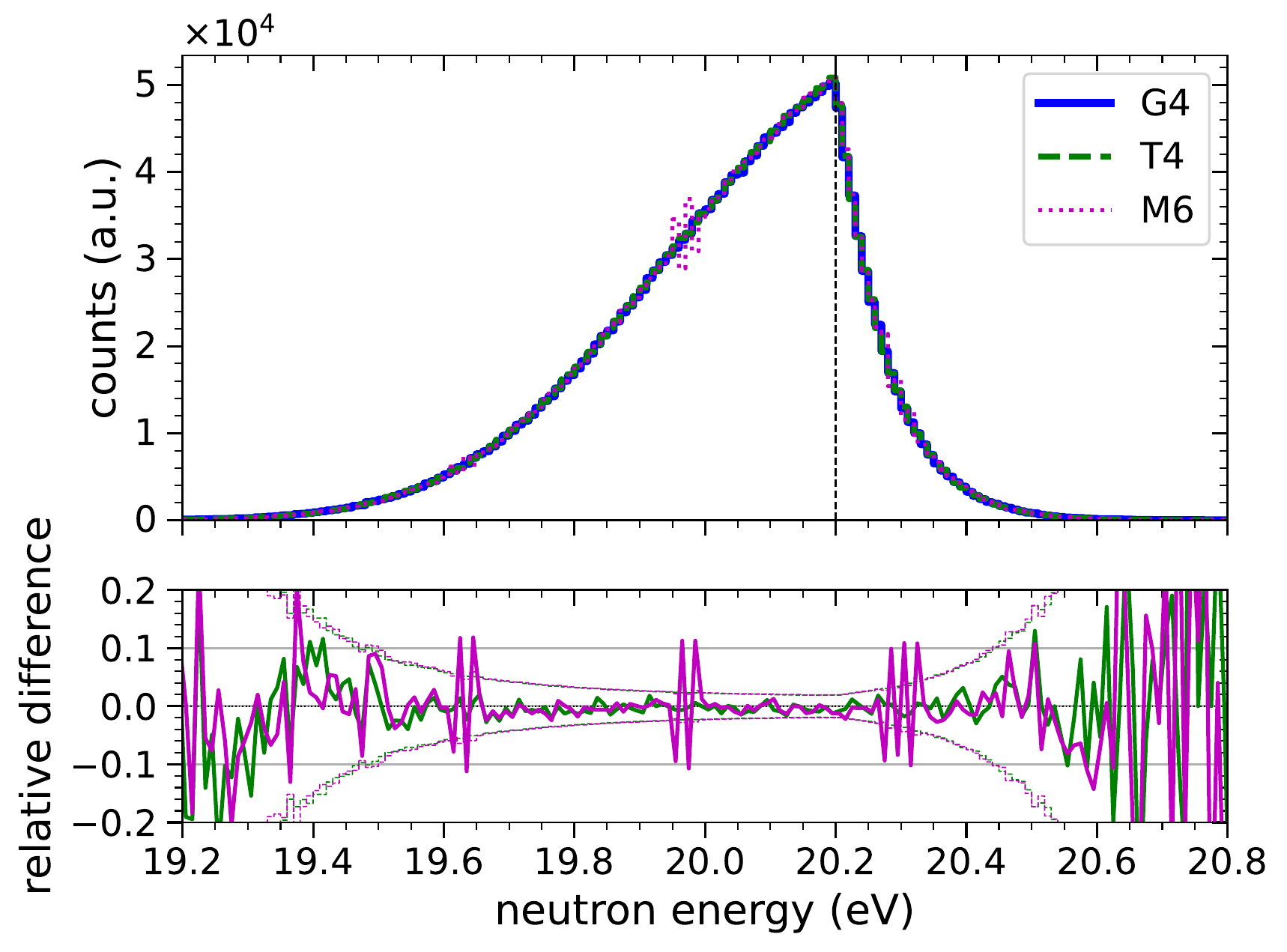}
         \caption{$20.2$ eV - $1000$ K - SVT}
     \end{subfigure}
     \hfill
     \begin{subfigure}[b]{0.49\textwidth}
         \centering
         \includegraphics[width=\textwidth]{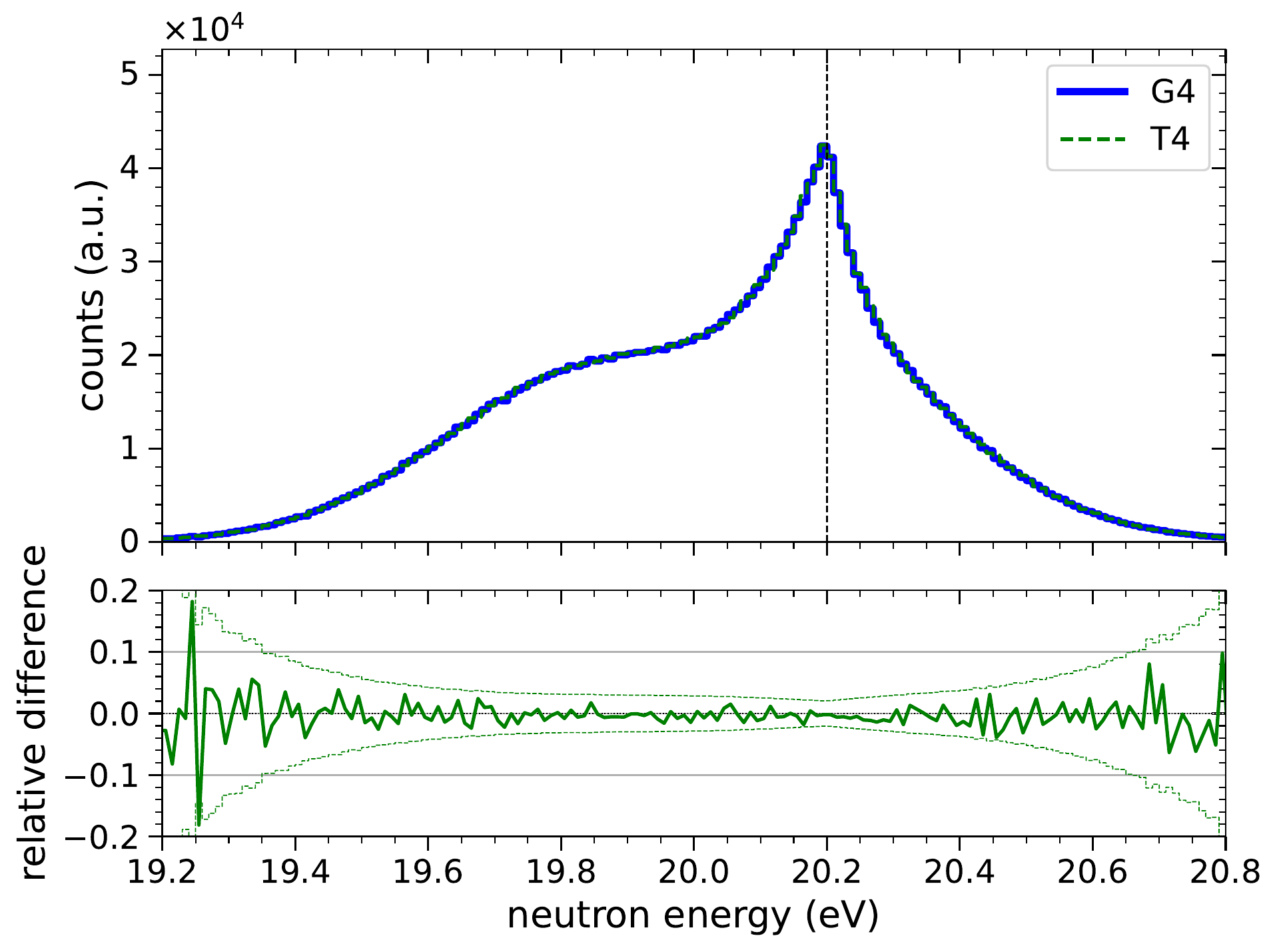}
         \caption{$20.2$ eV - $1000$ K - DBRC}
     \end{subfigure}
        \caption{Outgoing neutron energy distributions of scattered neutrons for $20.2$ eV neutrons on $^{238}$U at different temperatures computed with SVT (left column) and DBRC (right column) computed with \geant and \tripoli. The bottom part presents the relative difference G4/T4-1 (green solid line) and G4/M6-1 (magenta solid line) with three times the standard statistical error of this difference (dashed lines). \mcnp results at 300 K and 1000 K have been normalised to \tripoli ones, because the Doppler broadened cross sections at 300 K and 1000 K are effectively taken at 293.6 K and 900 K (pre-processed cross sections).}
        \label{fig:comparison2}
\end{figure}

\begin{figure}[htbp]
     \centering
     \begin{subfigure}[b]{0.49\textwidth}
         \centering
         \includegraphics[width=\textwidth]{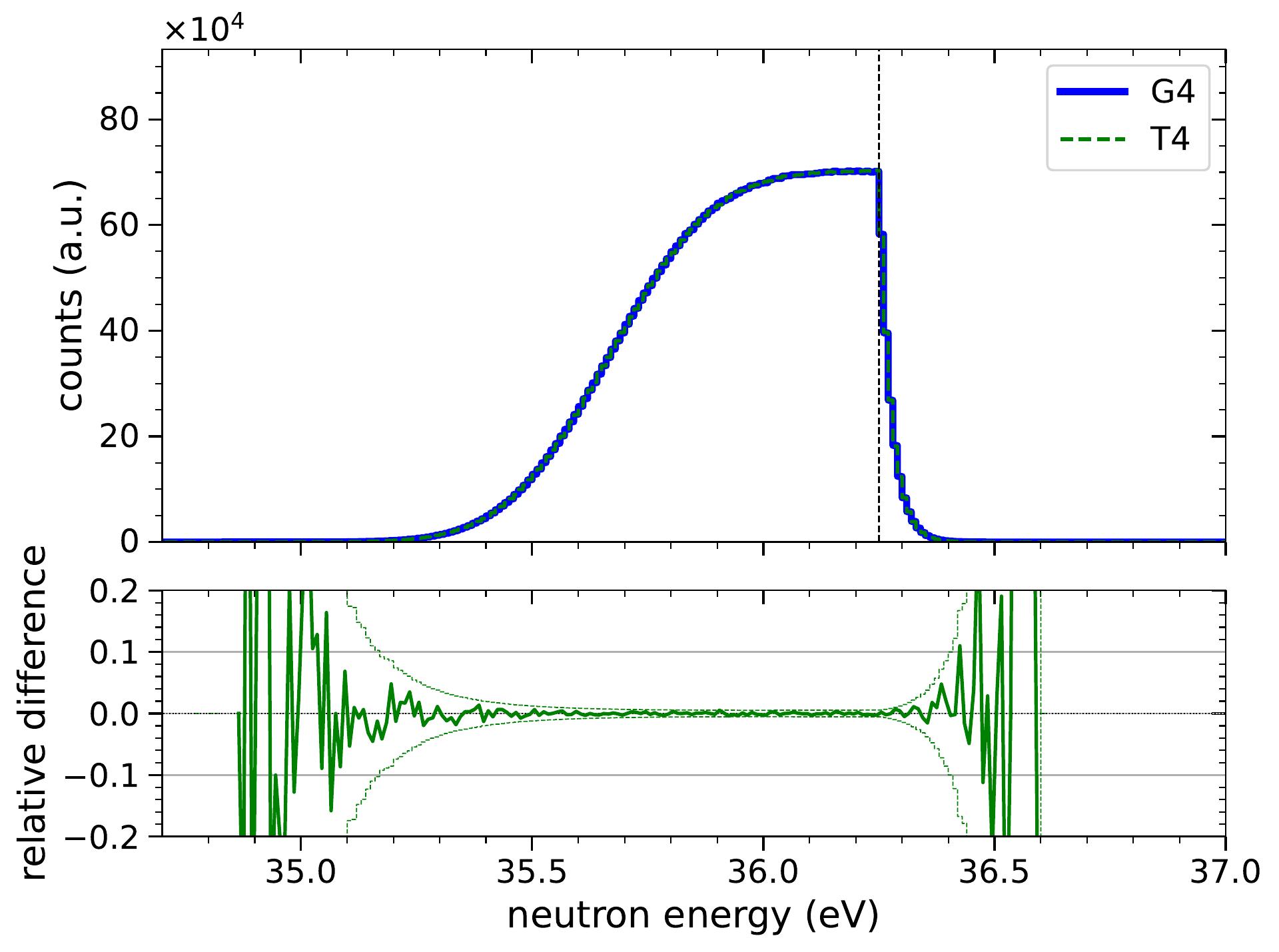}
         \caption{$36.25$ eV - $300$ K - SVT}
     \end{subfigure}
     \hfill
     \begin{subfigure}[b]{0.49\textwidth}
         \centering
         \includegraphics[width=\textwidth]{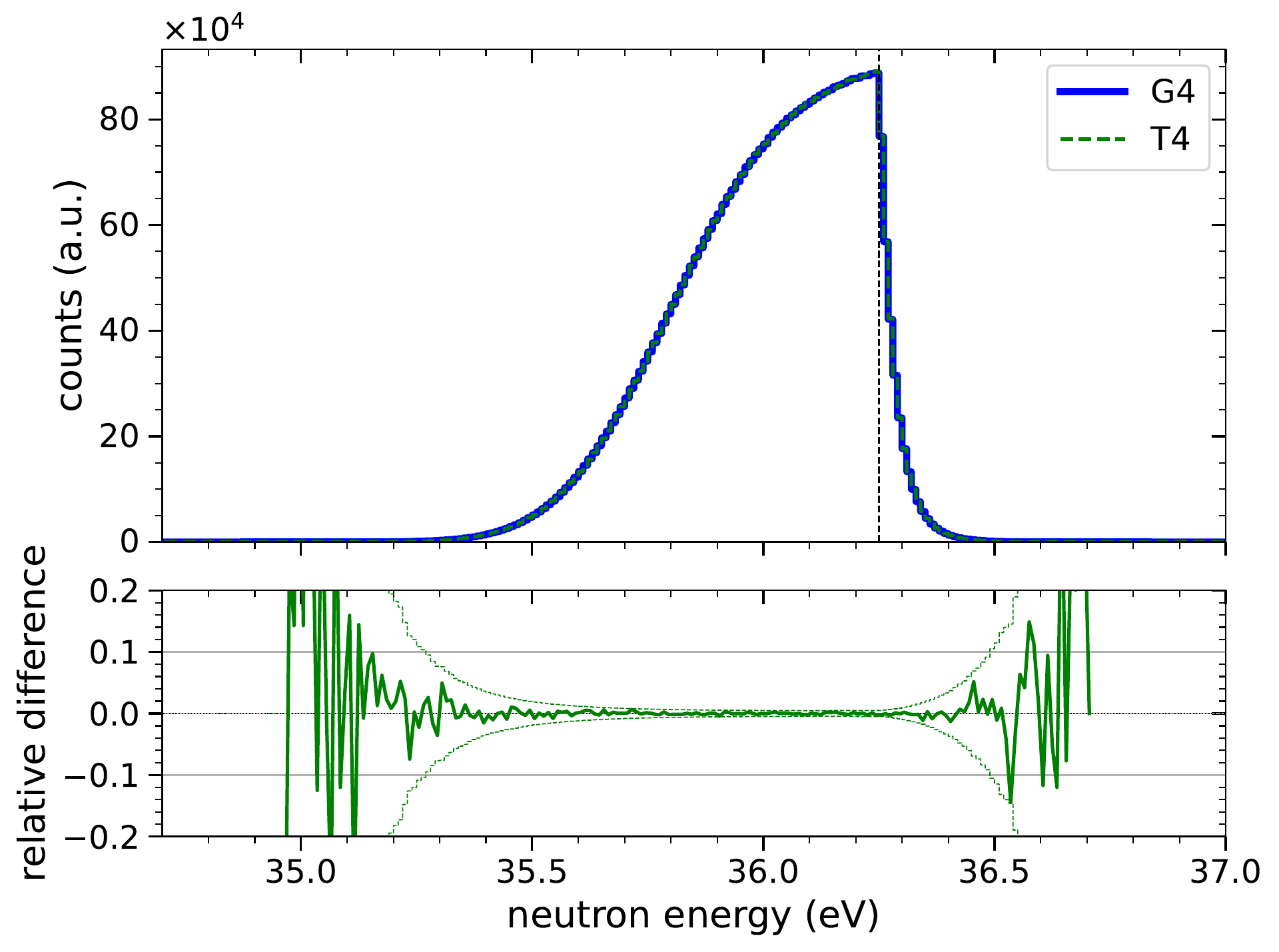}
         \caption{$36.25$ eV - $300$ K - DBRC}
     \end{subfigure}
     \begin{subfigure}[b]{0.49\textwidth}
         \centering
         \includegraphics[width=\textwidth]{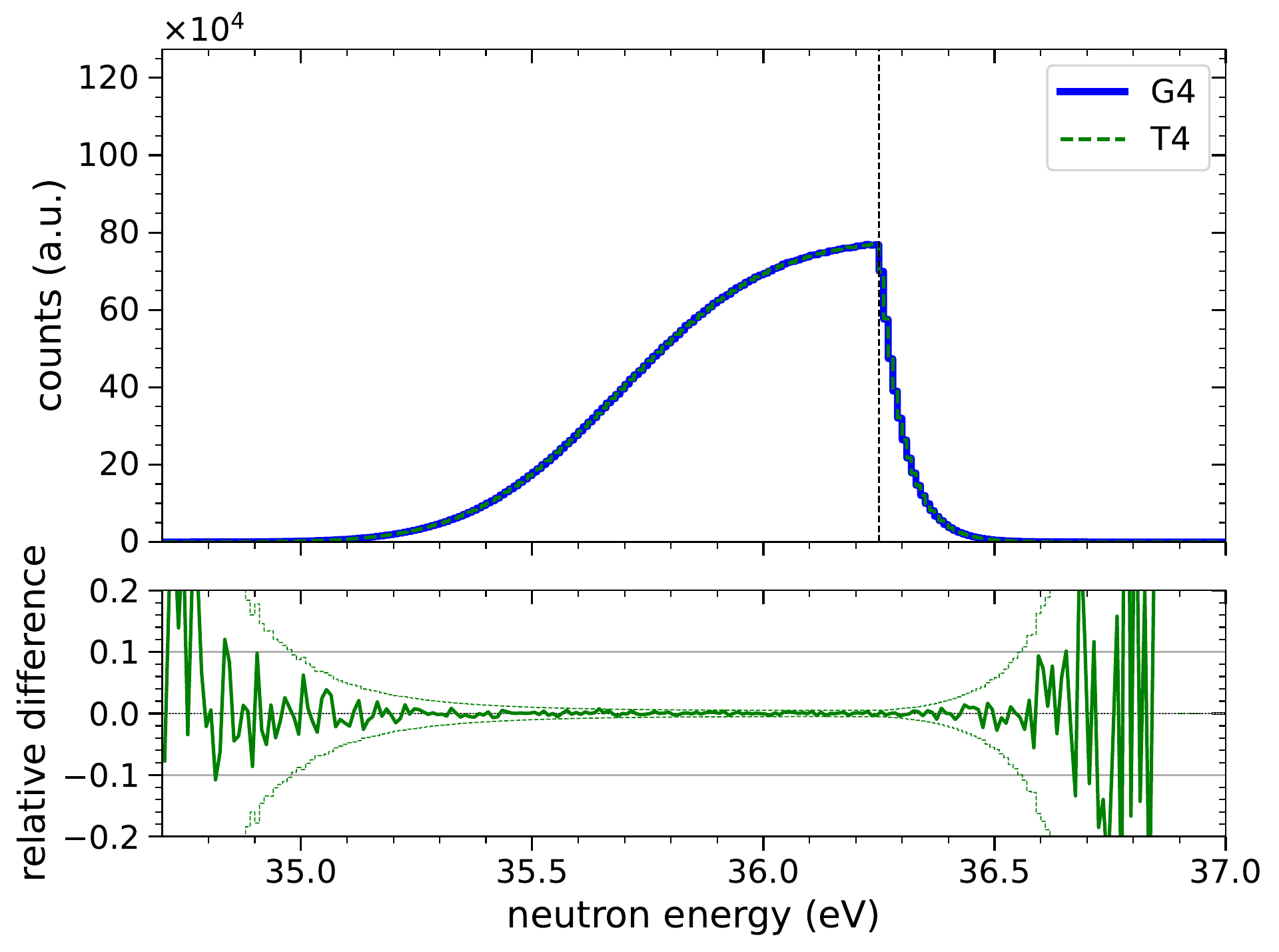}
         \caption{$36.25$ eV - $600$ K - SVT}
     \end{subfigure}
      \hfill
     \begin{subfigure}[b]{0.49\textwidth}
         \centering
         \includegraphics[width=\textwidth]{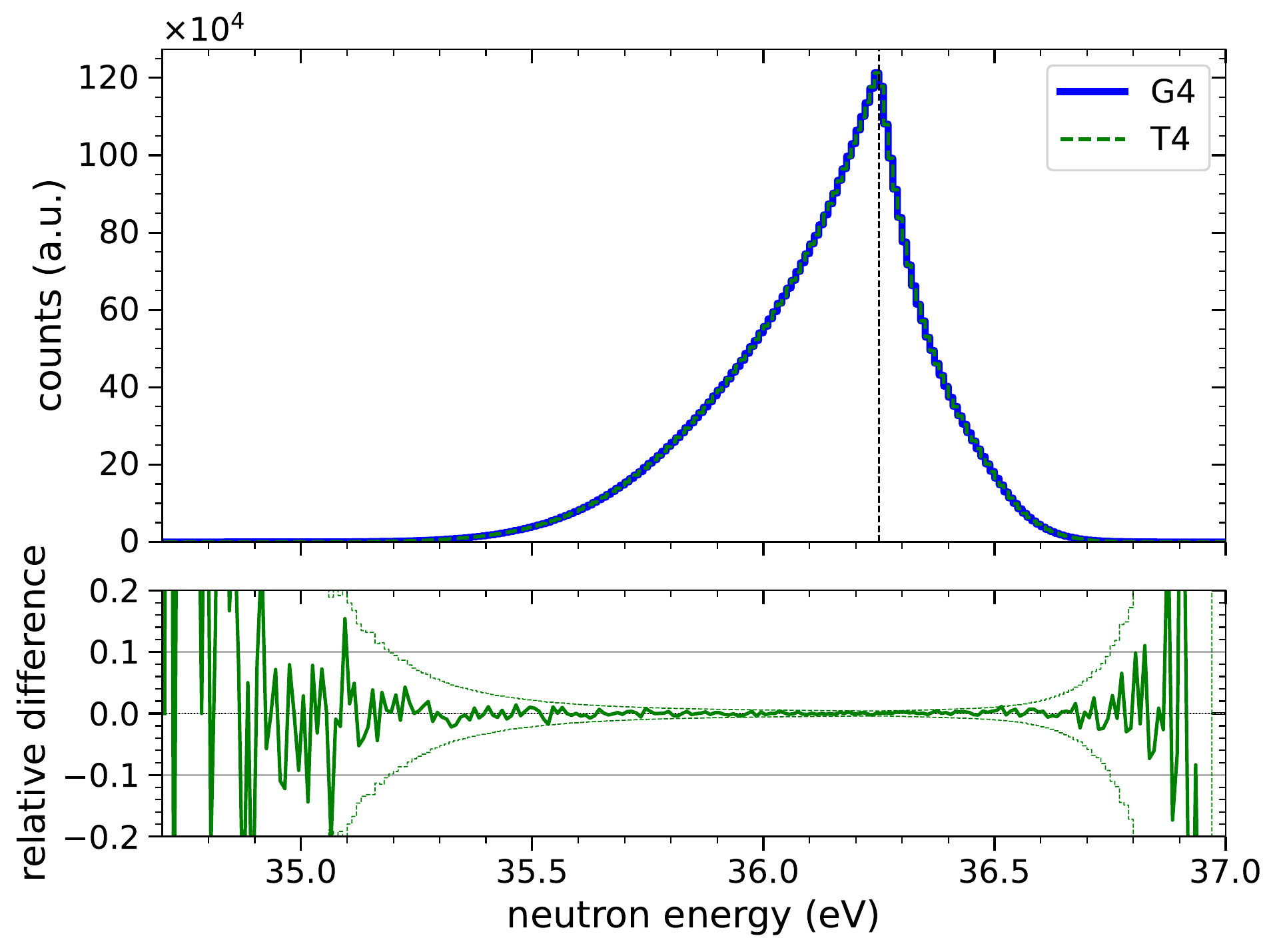}
         \caption{$36.25$ eV - $600$ K - DBRC}
     \end{subfigure}
     \begin{subfigure}[b]{0.49\textwidth}
         \centering
         \includegraphics[width=\textwidth]{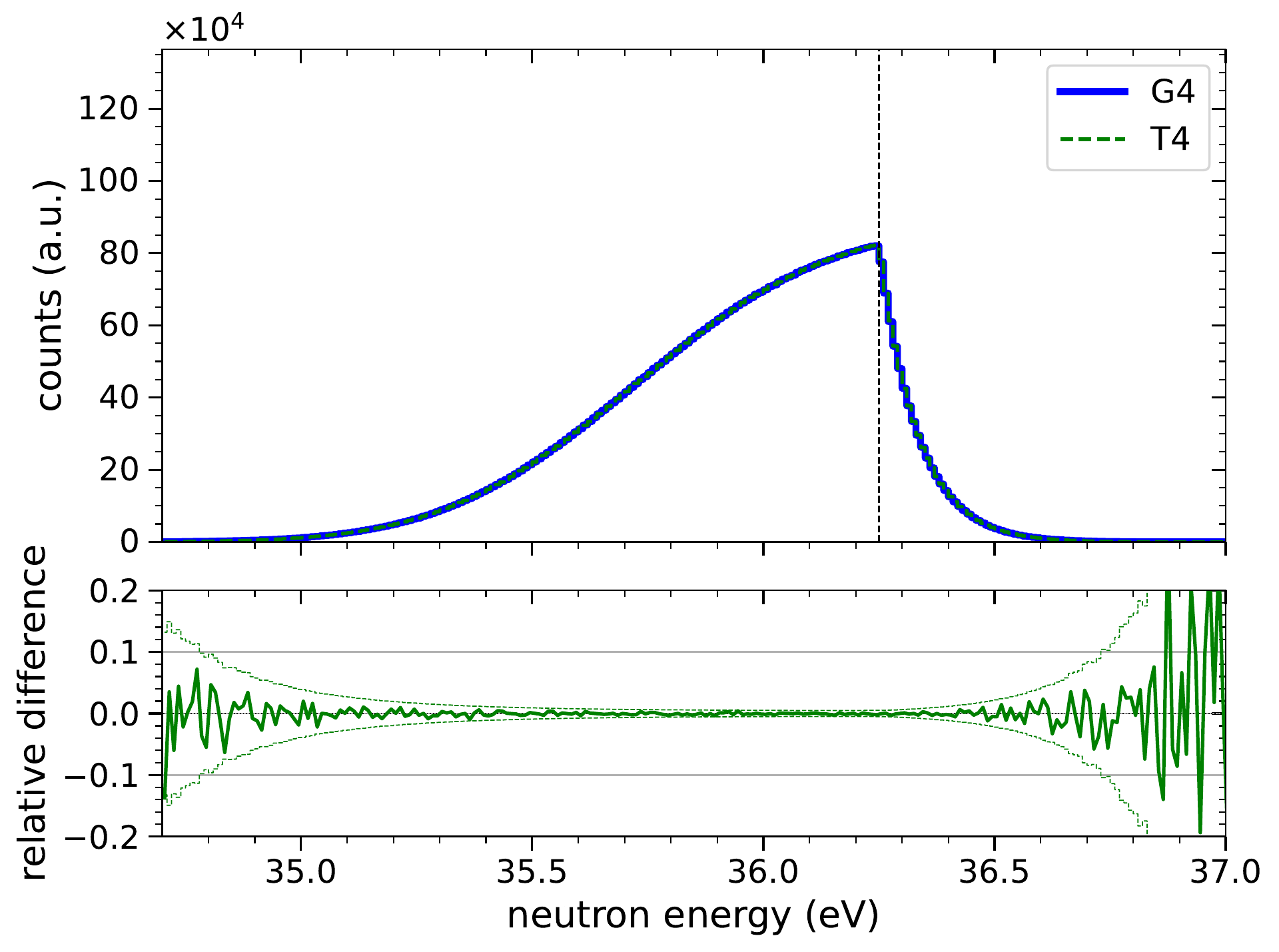}
         \caption{$36.25$ eV - $1000$ K - SVT}
     \end{subfigure}
     \hfill
     \begin{subfigure}[b]{0.49\textwidth}
         \centering
         \includegraphics[width=\textwidth]{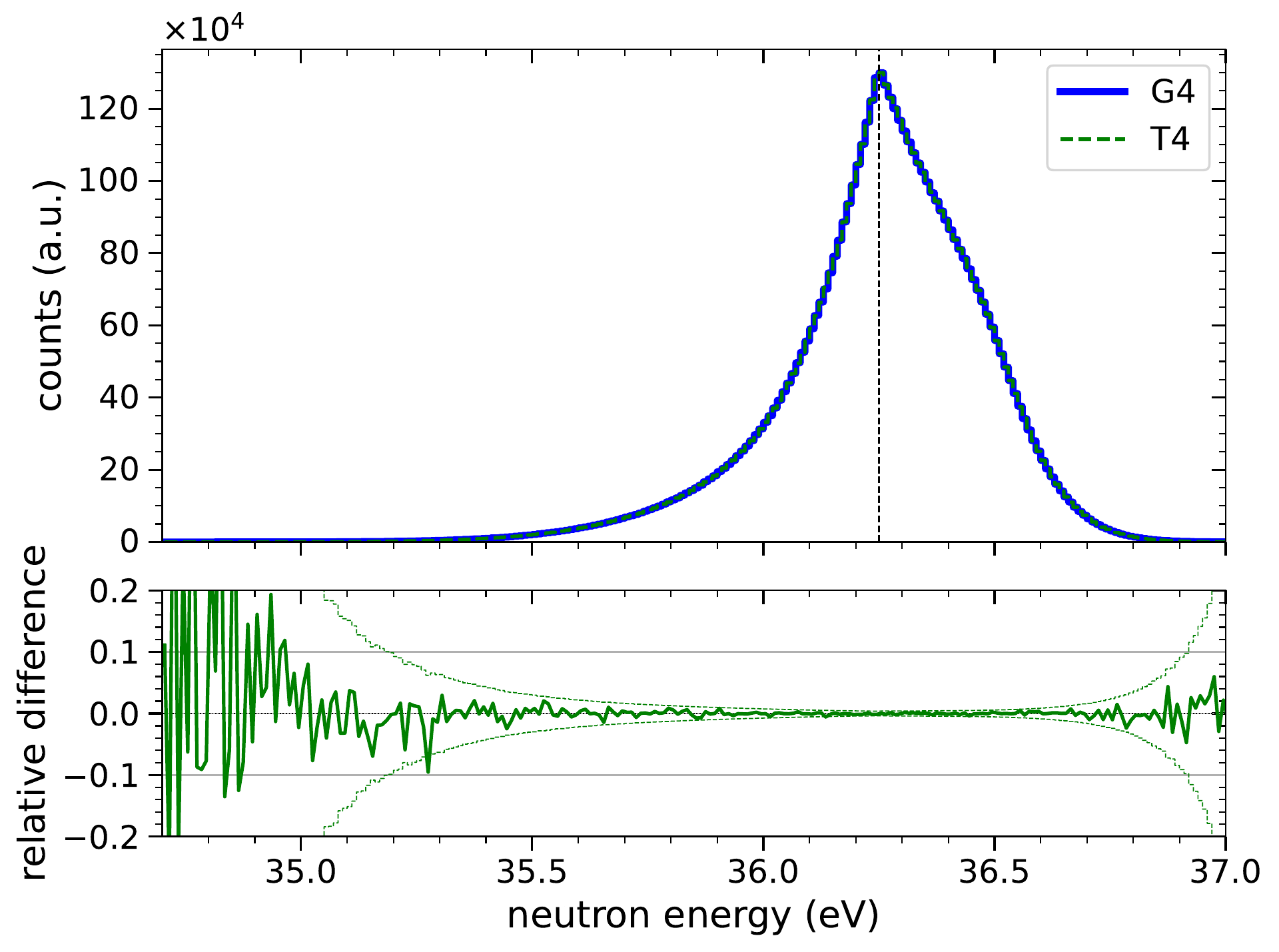}
         \caption{$36.25$ eV - $1000$ K - DBRC}
     \end{subfigure}
        \caption{Outgoing neutron energy distributions of scattered neutrons for $36.25$ eV neutrons on $^{238}$U at different temperatures computed with SVT (left column) and DBRC (right column) computed with \geant and \tripoli. The bottom part presents the relative difference G4/T4-1 (green solid line) with three times the standard statistical error of this difference (dashed line).}
        \label{fig:comparison3}
\end{figure}

The thin-cylinder benchmark has been performed in the same conditions as for the SVT benchmark. The DBRC results presented in the right columns of Figures \ref{fig:comparison1}, \ref{fig:comparison2} and \ref{fig:comparison3} show a very good agreement between \geant and \tripoli inside the statistical uncertainties. As can be seen, the secondary neutron energy distributions changes significantly compared to those obtained with the SVT method. Table \ref{tab:meanE}, presenting the average scattered neutron energy, shows that the average energy with the SVT method remains the same for the different temperatures while with the DBRC method it increases with the temperature. \geant and \tripoli agree well with each other since the relative errors are around $10^{-3}$~\% for both SVT and DBRC methods. The average neutron energy increase agrees with the increase of the up-scattering probability (\textit{i.e.} ratio of neutrons undergoing scattering with a higher energy than the initial neutron) as shown in Table~\ref{tab:upscattering}. Again, here, the results between \geant and \tripoli are in good agreement since the relative differences are less than $1 \%$. 
All these results validate the DBRC algorithm implementation in \geant.

\begin{table}[htbp]
\centering
\caption{Average energy of the scattered neutrons in eV for different primary energies and temperatures on $^{238}$U with SVT and DBRC methods computed with \geant and \tripoli. The relative difference (rel. diff.) between the two codes are in~$10^{-3}$ \%.}
\label{tab:meanE}
\begin{tabular}{rr|cc|cc|cc}
\multicolumn{1}{c}{E}    & \multicolumn{1}{c|}{T}   & \multicolumn{2}{c|}{\geant} & \multicolumn{2}{c|}{\tripoli} & \multicolumn{2}{c}{rel. diff.} \\
\multicolumn{1}{c}{(eV)} & \multicolumn{1}{c|}{(K)} & SVT          & DBRC        & SVT            & DBRC          & SVT            & DBRC          \\ \hline
6.52                     & 300                      & 6.466                   & 6.534                    & 6.466                   & 6.534                     & -0.85                   & -1.21                    \\
                         & 600                      & 6.466                   & 6.589                    & 6.466                   & 6.589                     & -0.47                   & -0.62                    \\
                         & 1000                     & 6.467                   & 6.605                    & 6.467                   & 6.605                     & -0.20                   & 0.08                     \\ \hline
20.2                     & 300                      & 20.031                  & 19.979                   & 20.031                  & 19.979                    & 0.57                    & -0.69                    \\
                         & 600                      & 20.031                  & 19.987                   & 20.031                  & 19.987                    & -0.22                   & 1.40                     \\
                         & 1000                     & 20.032                  & 20.043                   & 20.032                  & 20.044                    & 0.75                    & -1.96                    \\ \hline
36.25                    & 300                      & 35.946                  & 36.008                   & 35.946                  & 36.008                    & -0.02                   & -0.17                    \\
                         & 600                      & 35.946                  & 36.128                   & 35.946                  & 36.128                    & -0.07                   & 0.28                     \\
                         & 1000                     & 35.947                  & 36.256                   & 35.947                  & 36.256                    & -0.08                   & 0.02        
\end{tabular}
\end{table}

\begin{table}[htbp]
\centering
\caption{Up-scattering probability in \% for different primary energies and temperatures on $^{238}$U computed with SVT and DBRC methods with \geant and \tripoli. The relative difference (rel. diff.) between the two codes are in \%.}
\label{tab:upscattering}
\begin{tabular}{rr|cc|cc|cc}
\multicolumn{1}{c}{E}    & \multicolumn{1}{c|}{T}   & \multicolumn{2}{c|}{\geant} & \multicolumn{2}{c|}{\tripoli} & \multicolumn{2}{c}{rel. diff.} \\
\multicolumn{1}{c}{(eV)} & \multicolumn{1}{c|}{(K)} & SVT          & DBRC        & SVT            & DBRC          & SVT            & DBRC          \\ \hline
6.52                     & 300                      & 18.124                  & 62.294                   & 18.073                  & 62.250                    & 0.28                    & 0.07                     \\
                         & 600                      & 25.452                  & 82.798                   & 25.437                  & 82.832                    & 0.06                    & -0.04                    \\
                         & 1000                     & 30.407                  & 84.325                   & 30.365                  & 84.354                    & 0.14                    & -0.03                    \\ \hline
20.2                     & 300                      & 7.499                   & 5.692                    & 7.470                   & 5.681                     & 0.38                    & 0.20                     \\
                         & 600                      & 13.502                  & 15.456                   & 13.413                  & 15.434                    & 0.67                    & 0.14                     \\
                         & 1000                     & 18.848                  & 30.725                   & 18.789                  & 30.791                    & 0.31                    & -0.21                    \\ \hline
36.25                    & 300                      & 4.242                   & 7.135                    & 4.236                   & 7.143                     & 0.15                    & -0.10                    \\
                         & 600                      & 8.264                   & 30.658                   & 8.265                   & 30.629                    & -0.01                   & 0.10                     \\
                         & 1000                     & 12.712                  & 55.275                   & 12.713                  & 55.264                    & -0.01                   & 0.02               
\end{tabular}
\end{table}

\newpage
\section{On-the-fly cross section Doppler broadening in \geant}
\label{sec:DB_cross section}

\begin{figure}[H]
     \centering
     \begin{subfigure}[b]{0.49\textwidth}
         \centering
         \includegraphics[width=\textwidth]{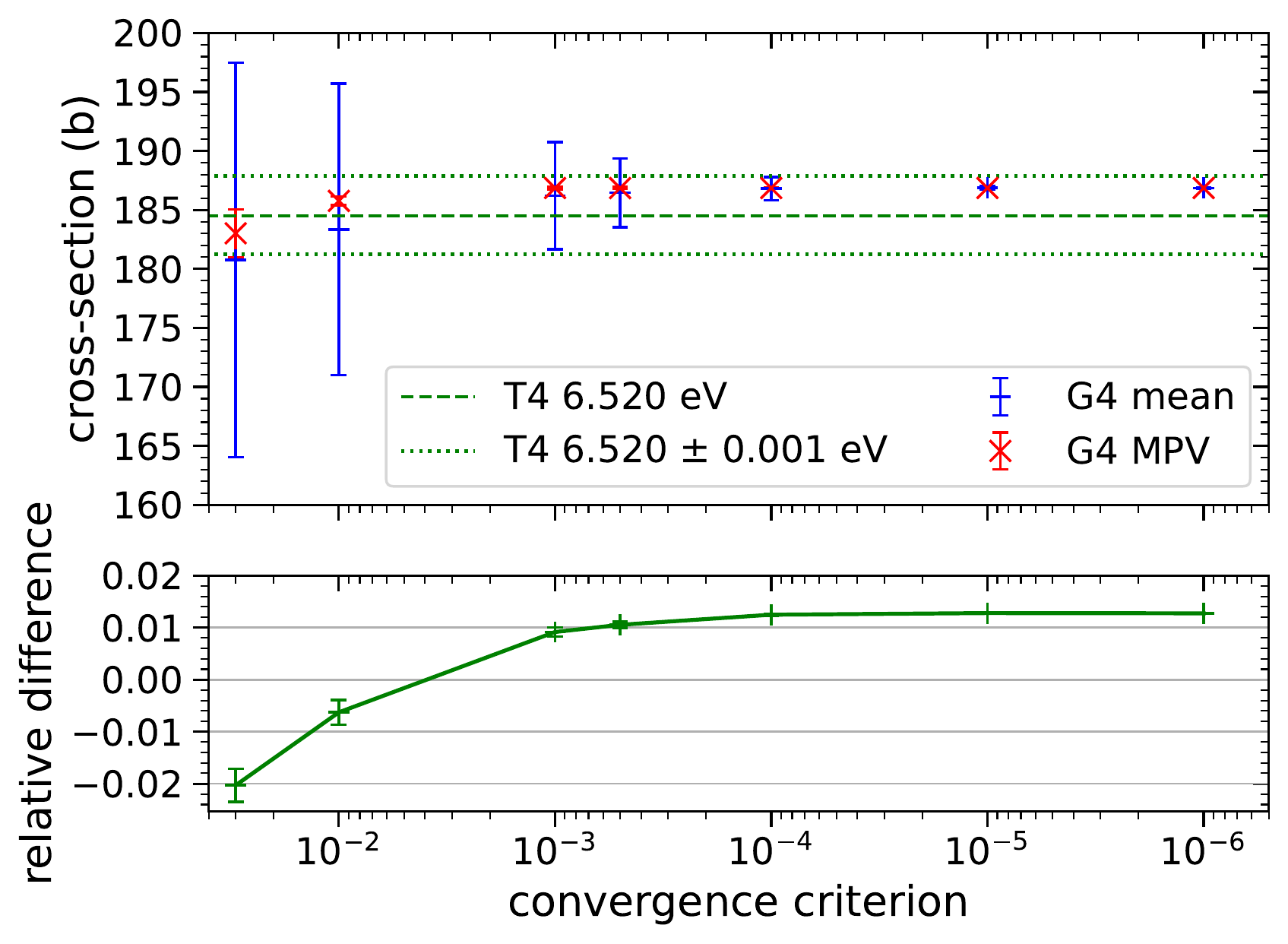}
         \caption{radiative capture}
         \label{fig:capturexs}
     \end{subfigure}
     \hfill
     \begin{subfigure}[b]{0.49\textwidth}
         \centering
         \includegraphics[width=\textwidth]{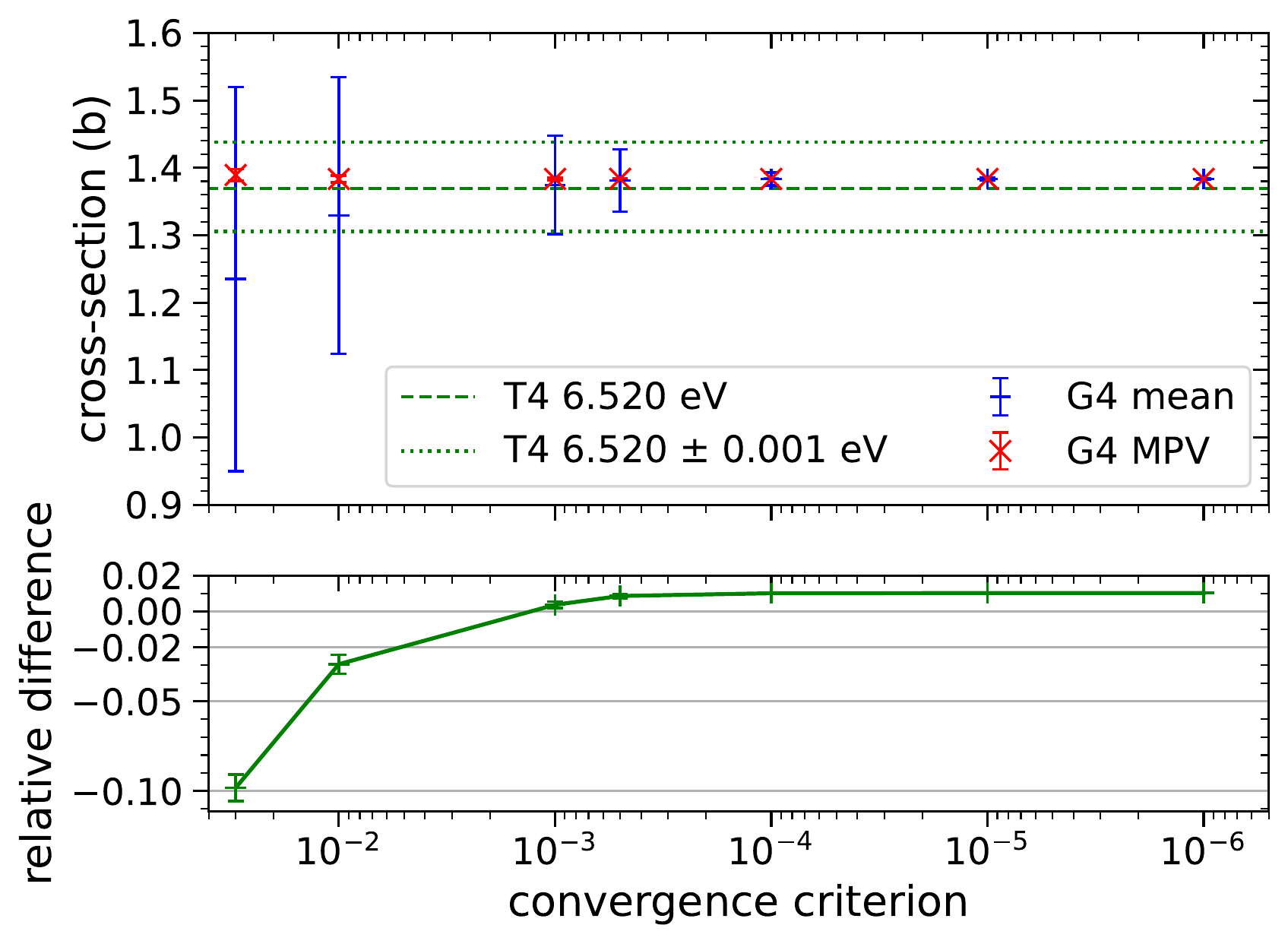}
         \caption{elastic scattering}
         \label{fig:elasticxs}
     \end{subfigure}
        \caption{Mean and most probable value (MPV), along with their standard deviations, of the $^{238}$U cross sections at $6.52$ eV and $300$ K as a function of the convergence criterion $\epsilon_\text{xs}$ calculated in \geant (G4). The cross section used in \tripoli (T4) is shown as green dashed line. The cross section relative difference \mbox{G4/T4-1} with its statistical error of the mean is shown at the bottom.}
        \label{fig:xs}
\end{figure}

\begin{figure}[H]
     \centering
     \begin{subfigure}[b]{0.49\textwidth}
         \centering
         \includegraphics[width=\textwidth]{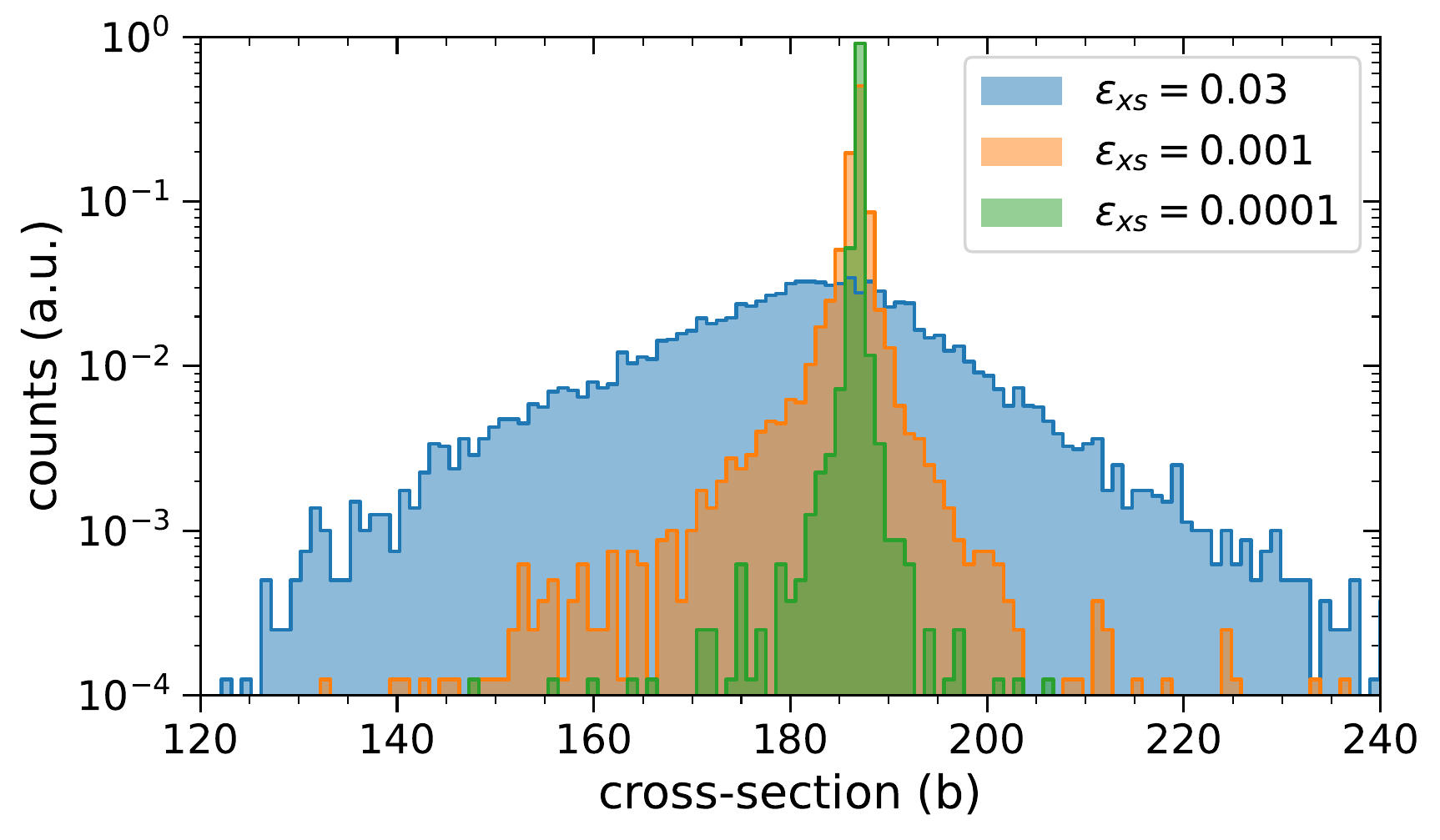}
         \caption{radiative capture}
         \label{fig:capturedistri}
     \end{subfigure}
     \hfill
     \begin{subfigure}[b]{0.49\textwidth}
         \centering
         \includegraphics[width=\textwidth]{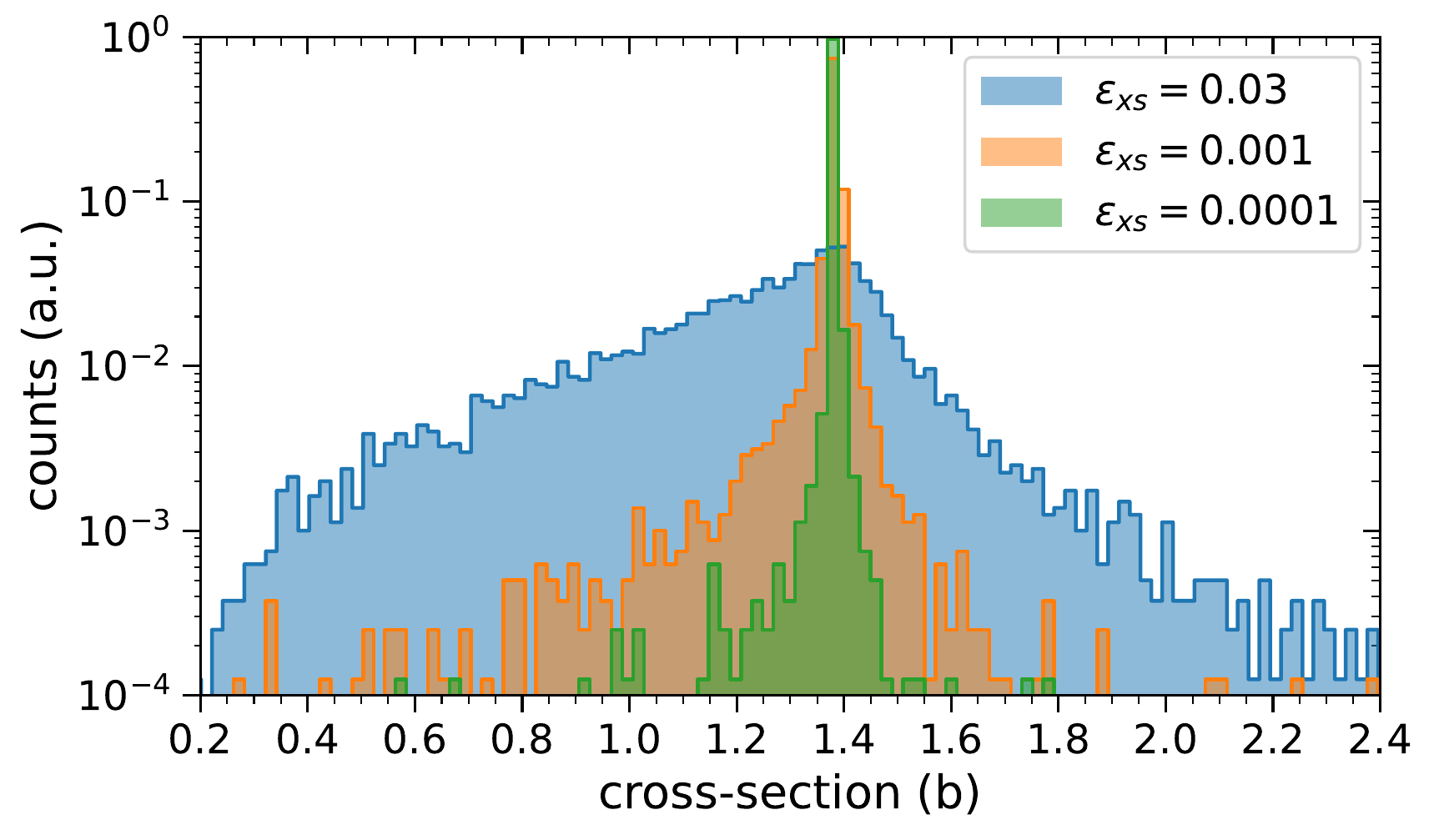}
         \caption{elastic scattering}
         \label{fig:elasticdistri}
     \end{subfigure}
        \caption{Normalised Doppler broadened cross section distributions for $^{238}$U for different convergence criteria $\epsilon_\text{xs}$.}
        \label{fig:distri}
\end{figure}

\begin{figure}[H]
    \centering
    \includegraphics[width=0.49\textwidth]{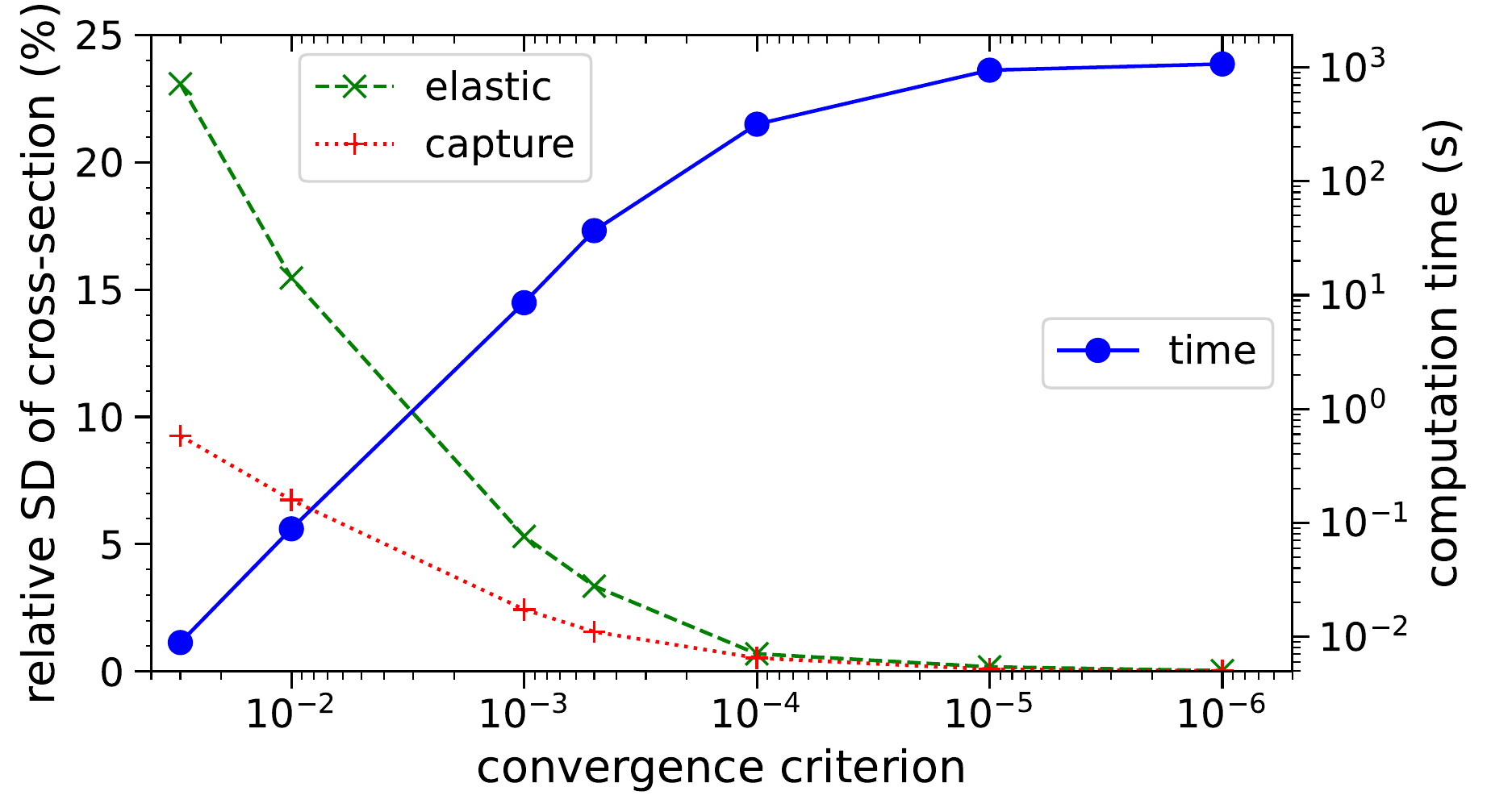}
    \caption{Relative standard deviation (SD) in $\%$ for elastic and capture $^{238}$U cross sections and computation time for different convergence criteria $\epsilon_\text{xs}$.}
    \label{fig:time}
\end{figure}

\begin{figure}[H]
     \centering
     \begin{subfigure}[b]{0.49\textwidth}
         \centering
         \includegraphics[width=\textwidth]{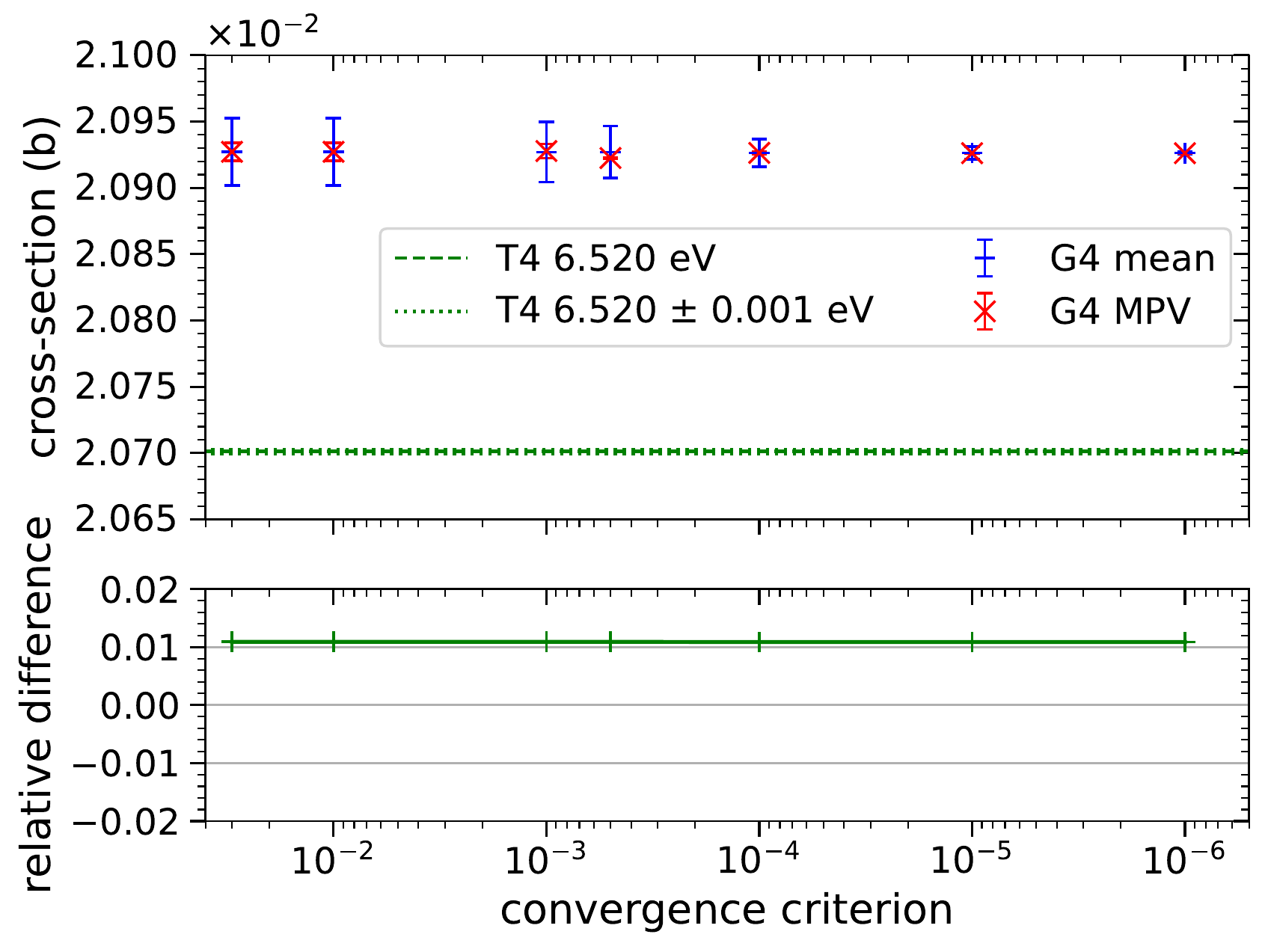}
         \caption{radiative capture}
         \label{fig:capturexsH}
     \end{subfigure}
     \hfill
     \begin{subfigure}[b]{0.49\textwidth}
         \centering
         \includegraphics[width=\textwidth]{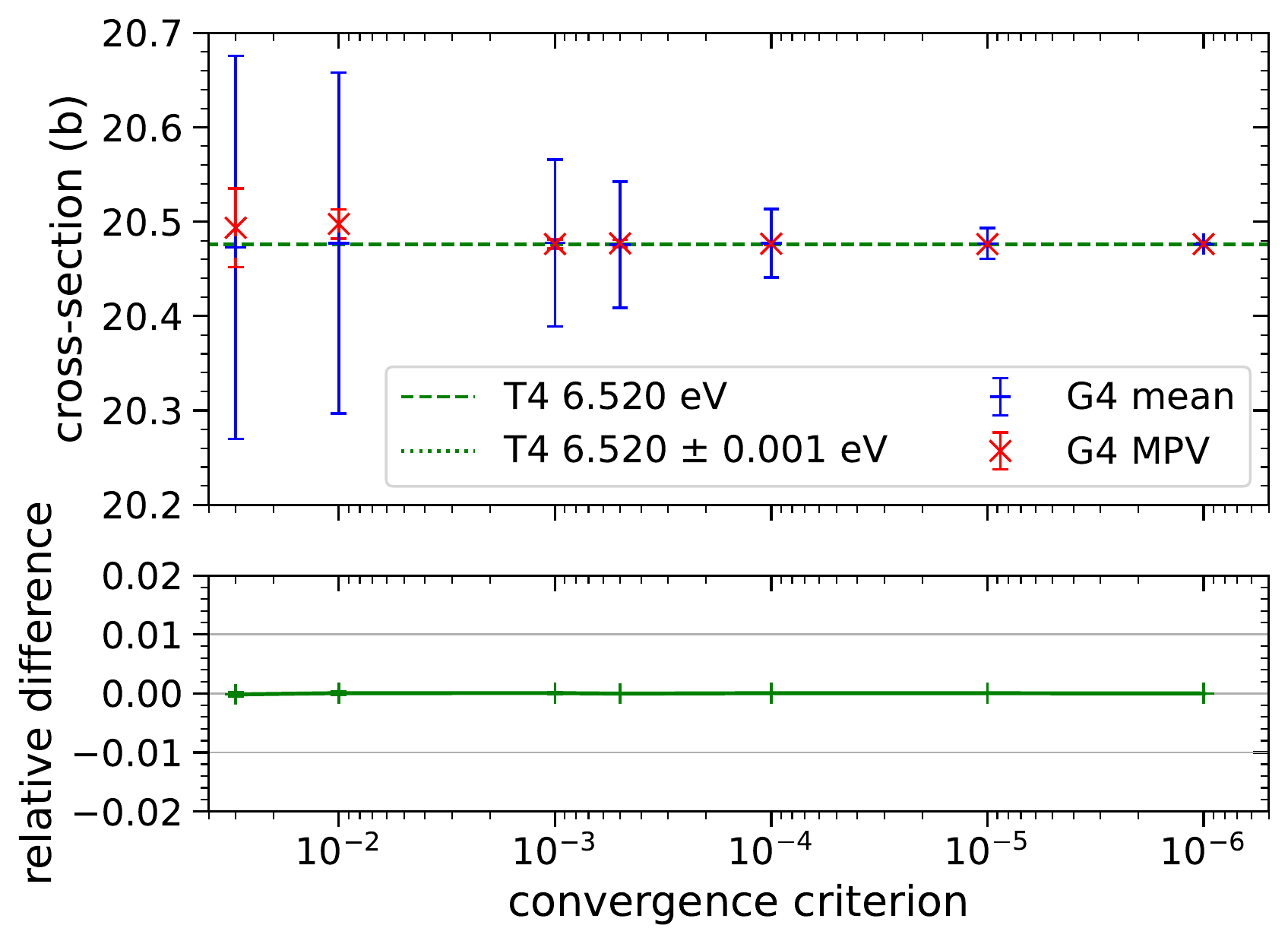}
         \caption{elastic scattering}
         \label{fig:elasticxsH}
     \end{subfigure}
        \caption{Mean and most probable value (MPV), along with their standard deviations, of the $^{1}$H cross sections at $6.52$ eV and $300$ K as a function of the convergence criterion $\epsilon_\text{xs}$ calculated in \geant (G4). The cross section used in \tripoli (T4) is shown as green dashed line. The cross section relative difference \mbox{G4/T4-1} with its statistical error of the mean is shown at the bottom.}
        \label{fig:xsH}
\end{figure}

\begin{figure}[H]
     \centering
     \begin{subfigure}[b]{0.49\textwidth}
         \centering
         \includegraphics[width=\textwidth]{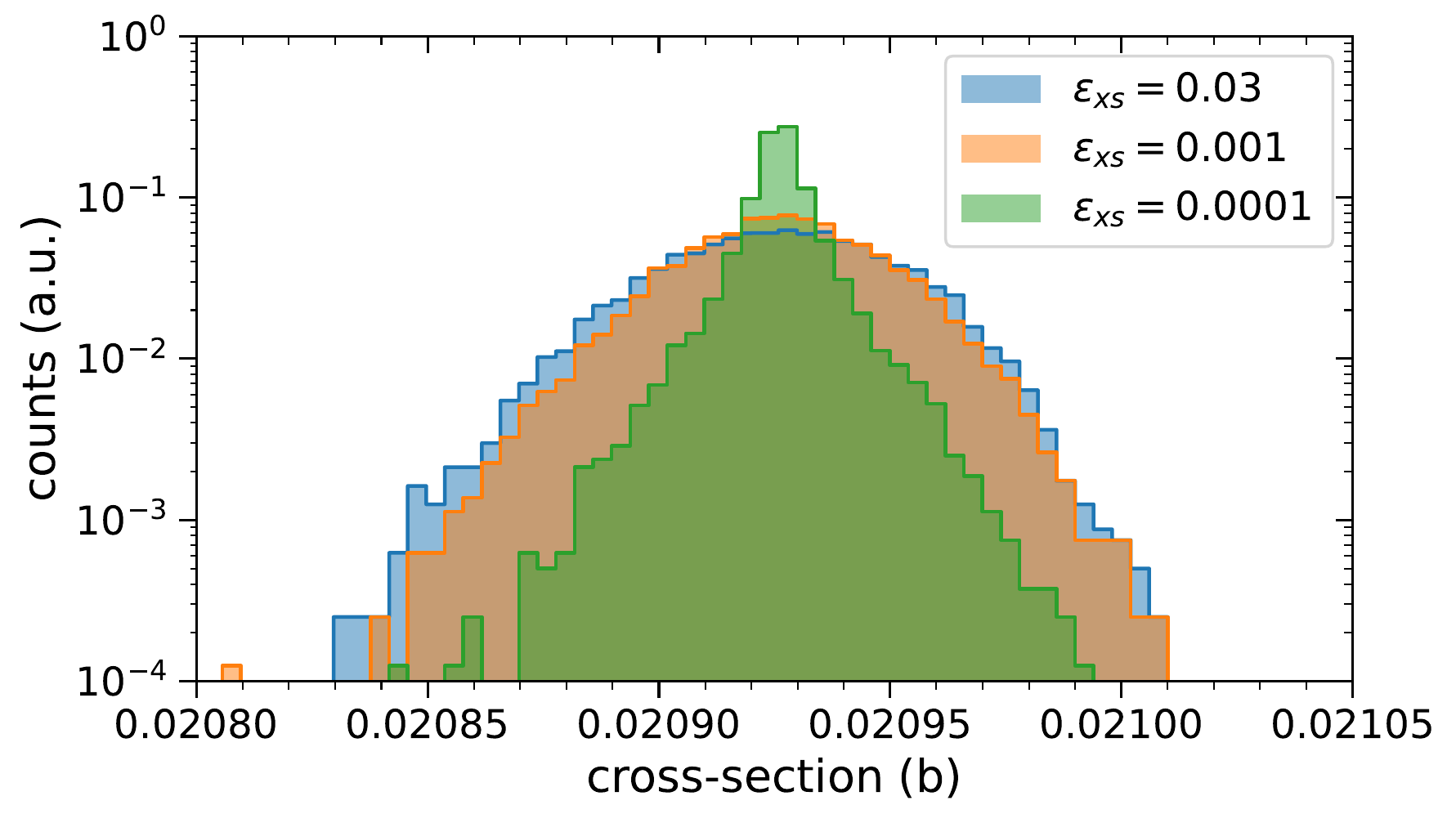}
         \caption{radiative capture}
         \label{fig:capturedistriH}
     \end{subfigure}
     \hfill
     \begin{subfigure}[b]{0.49\textwidth}
         \centering
         \includegraphics[width=\textwidth]{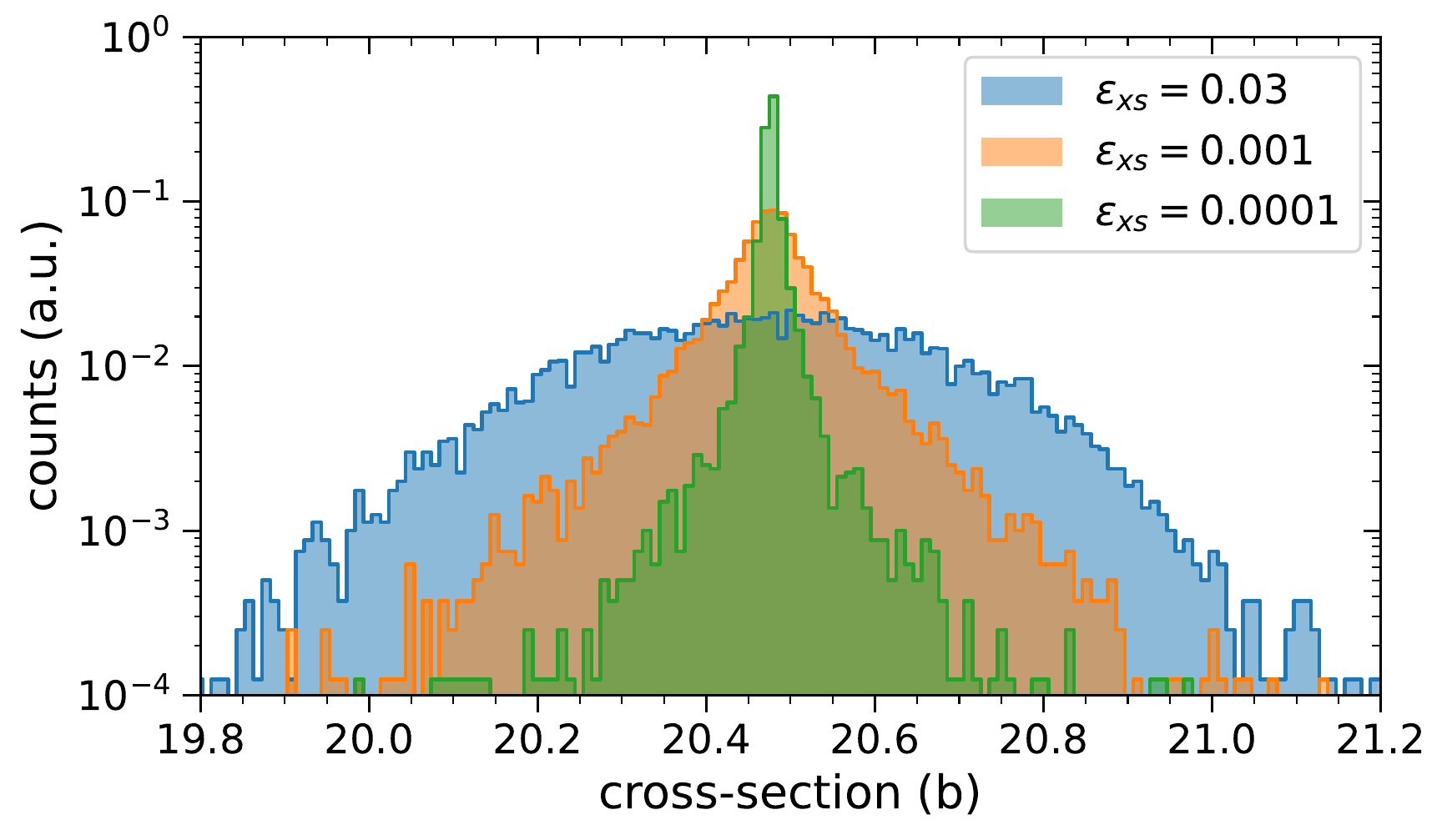}
         \caption{elastic scattering}
         \label{fig:elasticdistriH}
     \end{subfigure}
        \caption{Normalised Doppler broadened cross section distributions for $^{1}$H for different convergence criteria $\epsilon_\text{xs}$.}
        \label{fig:distriH}
\end{figure}

\begin{figure}[H]
    \centering
    \includegraphics[width=0.5\textwidth]{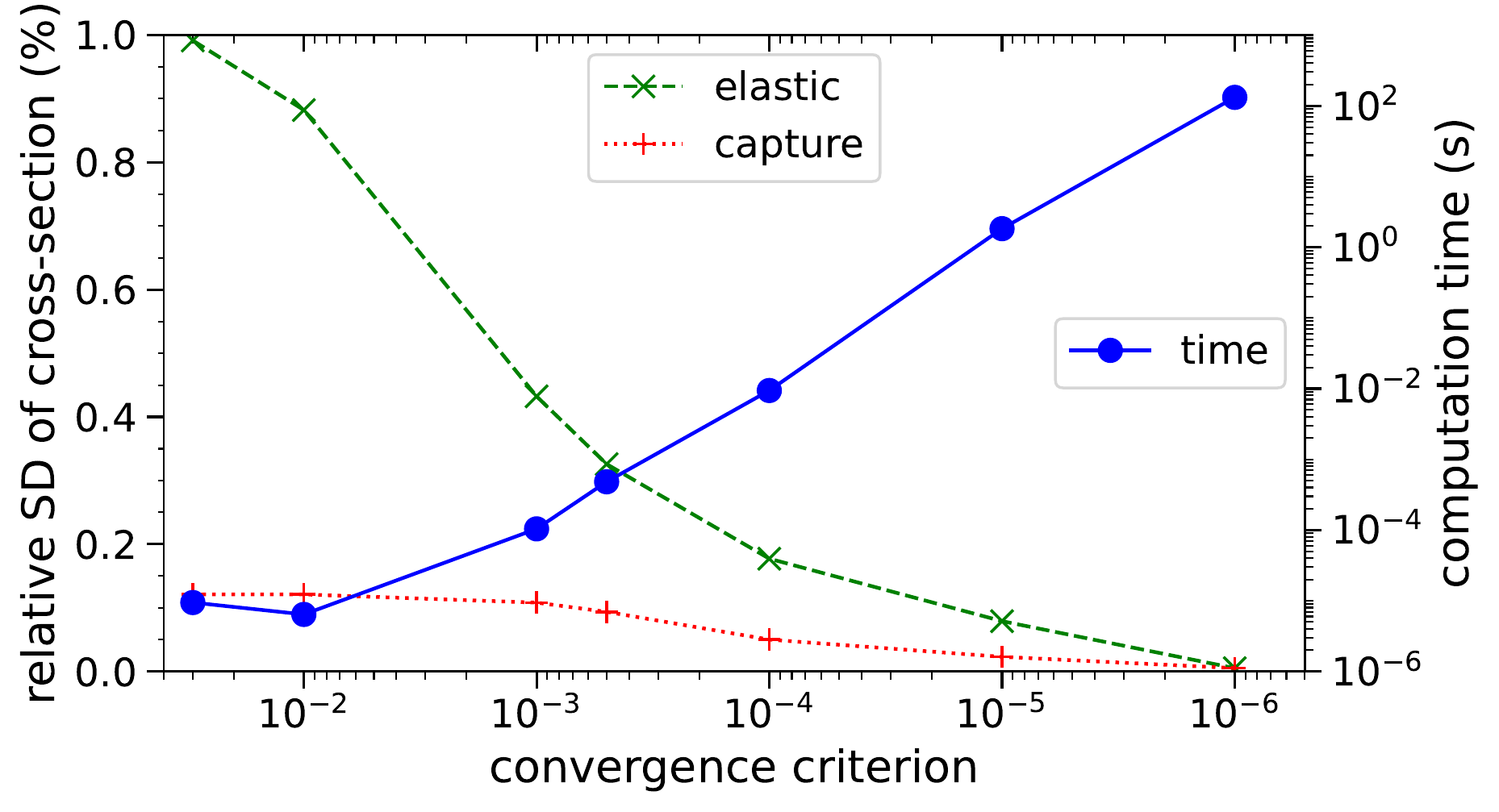}
    \caption{Relative standard deviation (SD) in $\%$ for elastic and capture $^{1}$H cross sections and computation time for different convergence criteria $\epsilon_\text{xs}$.}
    \label{fig:timeH}
\end{figure}

In Monte Carlo neutron transport codes, cross sections can be Doppler broadened mainly with two methods. The first method consists in performing exactly the Doppler broadening (given by Equation~\ref{eq:doppler}) before the simulation for example with the \mbox{SIGMA-1} algorithm implemented in a pre-processing tool such as \njoy \cite{Macfarlane2017}. Therefore no additional computation time is spent during the simulation for this. The drawbacks of this approach lies in the limited number of available temperatures for the simulation and the need to have a large computer memory size to store these data. It is mainly used in reference neutron transport codes such as \tripoli and \mcnp. The second method is more demanding in terms of computational time since the Doppler broadening is done during the simulation, \textit{i.e.} on-the-fly (OTF), but can be made at any temperature required by the user. This latter has been chosen in \geant for its versatility allowing to suit any needs from users coming from very different physics communities.
\newline
\indent
In comparing the thin-cylinder benchmark results from \geant and \tripoli, it was found that the total reaction rate from \geant was different from \tripoli, while relative values (such as the up-scattering probability) agree well between the two codes. This discrepancy arises because of the stochastic OTF Doppler broadening method implemented in \geant. Its implementation can be seen, for example, in the G4ParticleHPElasticData::GetCrossSection method for elastic scattering and can be summarised as followed:

\begin{lstlisting}[mathescape=true, language=C++, basicstyle=\small,label={lst:DBGeant4_OTF}, caption={Pseudo-code of the implemented on-the-fly stochastic integral of the Doppler broadened cross section in \geant.} ]

///Initialization
$size=\text{max}(10, T/60)$

///On-the-fly stochastic integral computation
while ($\left|\overline{\sigma_T^\text{old}(v_n)}-\overline{\sigma_T^\text{new}(v_n)}\right| > \epsilon_\text{xs} \cdot \overline{\sigma_T^\text{old}(v_n)}$){
    $\overline{\sigma_T^\text{old}(v_n)}=\overline{\sigma_T^\text{new}(v_n)}$
    while ($counter<size$){
        sample $\mathbf{V_t}$ from Maxwell-Boltzmann distribution
        get $\sigma(v_r)$
        $\sigma_T^\text{new}(v_n)\pluseq\sigma(v_r)\cdot \frac{|\mathbf{V_t}-\mathbf{v_n}|}{v_n}$
        $counter\plusplus$
    }
    $\overline{\sigma_T^\text{new}(v_n)}=\frac{\sigma_T^\text{new}(v_n)}{counter}$
    $size\pluseq size$
}
return $\overline{\sigma_T^\text{new}(v_n)}$
\end{lstlisting}
\noindent
with $counter$ the total number of loops performed by the algorithm and $size$ the number of loops for each batch. At each step, the Doppler broadened cross section is computed and compared with its previous value. The algorithm stops when a convergence criterion $\epsilon_\text{xs}$ is met. 
\newline
\indent
By default $\epsilon_\text{xs}$ are set to 0.03 and 0.01 respectively for capture and elastic reactions and for fission and inelastic reactions. However, these criteria do not guarantee sufficient cross section convergence as already pointed out in \cite{Cai2014} and as can be seen in Figures~\ref{fig:capturexs}~and~\ref{fig:elasticxs}. Therefore the impact of $\epsilon_\text{xs}$ on the capture and elastic scattering has been studied with the thin-cylinder benchmark at 300~K and with 6.52~eV neutrons. The quantities of interest are the mean cross section, its standard deviation and its most probable value (MPV) since the distributions are not symmetric as can be seen in Figures~\ref{fig:capturedistri}~and~\ref{fig:elasticdistri}. The \geant OTF algorithm has been performed 8000 times to get statistical significant results for each convergence criterion. The results are compared to \tripoli cross sections which have been computed exactly in using the \mbox{SIGMA-1} algorithm with \njoy.
\newline
\indent
The Figures~\ref{fig:capturexs}~and~\ref{fig:elasticxs}, presenting the mean and MPV cross sections, show that the mean cross section converges when $\epsilon_\text{xs}$ decreases and that the MPV is always approximately the same for all convergence criterion values and is equal to the converged mean cross section value. 
With the default value $\epsilon_\text{xs}$=0.03, the mean capture and elastic cross sections are respectively around $2 \%$ and $10 \%$ lower than \tripoli ones. For $\epsilon_\text{xs}<$0.001, the elastic and capture mean cross sections are converged to a value $1 \%$ higher than \tripoli ones as shown in Figures~\ref{fig:capturexs} and \ref{fig:elasticxs}. This $1 \%$ discrepancy could be caused by interpolation errors since \geant results are compatible with \tripoli cross sections taken at $6.520 \text{eV} \pm 1$~meV. This discrepancy is therefore not investigated further.
\newline 
\indent
The evolution of the cross section standard deviation is also quantified in Figure \ref{fig:time} as a function of $\epsilon_\text{xs}$ and shows a decrease from 23 $\%$ for elastic cross section ($9 \%$ for capture) at $\epsilon_\text{xs}=0.03$ to less than~$1 \%$ for both cross sections for convergence criterion at $\epsilon_\text{xs}=10^{-4}$. With these results, one can be tempted to change the default $\epsilon_\text{xs}$ value to a lower one, however a lower convergence criterion leads to a higher number of loops and therefore a higher computation time. In fact the computation time rises drastically as a function of $\epsilon_\text{xs}$ and then its increase starts to slow down from $\epsilon_\text{xs}=10^{-4}$. In the future it could be worth to evaluate how this OTF algorithm performs compared to other algorithms such as the regression method \cite{Yesilyurt2012,Martin2014}, the multipole representation of the cross section \cite{Forget2014} or methods based on explicit treatment of target motion at collision sites \cite{Viitanen2012}.
\newline 
\indent
It has to be pointed out that all the results presented in Section \ref{sec:DBRC} have been computed with a convergence criterion set to $\epsilon_\text{xs}=10^{-4}$ which is a trade-off between the cross section precision and computation time. Unfortunately this only can be done for benchmarking purposes, because of the prohibitive computation time for one OTF Doppler broadening algorithm call. However this problem would not be so important for integral studies since the Doppler broadening will be performed many times (at different energies) and so in average will give the right result.
\newline
\indent
It finally has to be recalled that this behaviour arises because the resonant part of the cross section is investigated. In fact for non-resonant cross sections, such as for $^{1}$H, Figure \ref{fig:xsH} shows that the cross section mean and MPV are approximately the same for all convergence criteria and agree with \tripoli cross sections. The capture cross section is again around $1$ \% higher, which could be caused by interpolation errors because the $0$ K cross section for elastic scattering is almost constant at $6.52$ eV, whereas the capture cross section significantly decreases. The standard deviation is under $1$ \% for elastic and $0.2$ \% for capture for the default convergence criteria as can be seen in Figures \ref{fig:distriH} and \ref{fig:timeH}. 

\clearpage
\newpage

\newpage
\clearpage

\newpage

\section{Conclusion}
The Doppler Broadening Rejection Correction algorithm used to describe accurately the elastic scattering kernel of resonant heavy nuclei has been successfully implemented into \geant. The validation performed with the reference Monte Carlo neutron transport code \tripoli showed a very good agreement between the two codes within the statistical uncertainties. The DBRC algorithm will be implemented in the next \geant release. This work makes the Neutron-HP package at the state-of-the-art regarding the description of Doppler broadened elastic scattering kernels. The importance of the convergence criterion for the on-the-fly Doppler broadening routine has also been investigated and showed that the default convergence criterion values should be adapted on a simulation-by-simulation basis depending on the observable under study. With the developments performed in this work, the last missing feature in \geant compared to reference neutron transport codes is the description of the unresolved resonance region with probability tables. This last point will be addressed soon in a dedicated paper.

\section*{Acknowledgement}

This work could be done thanks to the IAEA´s technical cooperation programme, which supported a fellowship of Marek Zmeskal at IRFU, CEA in the framework of national project CZR0011 - Strengthening Human Resources Capacity in Nuclear Science and Technology.
The authors wish to sincerely thank Cédric Jouanne and Alberto Ribon head of the \geant Hadronic Working Group for fruitful discussions respectively regarding \tripoli and \geant . 
\tripoli~is a registered trademark of CEA.

\appendix
\bibliographystyle{unsrtnat}
\bibliography{references}

\begin{thebibliography}{35}
\providecommand{\natexlab}[1]{#1}
\providecommand{\url}[1]{\texttt{#1}}
\expandafter\ifx\csname urlstyle\endcsname\relax
  \providecommand{\doi}[1]{doi: #1}\else
  \providecommand{\doi}{doi: \begingroup \urlstyle{rm}\Url}\fi

\bibitem[et~al.(2003)]{Agostinelli2003}
S.~Agostinelli et~al.
\newblock {Geant4—a simulation toolkit}.
\newblock \emph{Nuclear Instruments and Methods in Physics Research Section A:
  Accelerators, Spectrometers, Detectors and Associated Equipment},
  506:\penalty0 250--303, 7 2003.
\newblock ISSN 01689002.
\newblock \doi{10.1016/S0168-9002(03)01368-8}.

\bibitem[et~al.(2006)]{Allison2006}
J.~Allison et~al.
\newblock {Geant4 developments and applications}.
\newblock \emph{IEEE Transactions on Nuclear Science}, 53:\penalty0 270--278, 2
  2006.
\newblock ISSN 0018-9499.
\newblock \doi{10.1109/TNS.2006.869826}.

\bibitem[et~al.(2016)]{Allison2016}
J.~Allison et~al.
\newblock {Recent developments in Geant4}.
\newblock \emph{Nuclear Instruments and Methods in Physics Research Section A:
  Accelerators, Spectrometers, Detectors and Associated Equipment},
  835:\penalty0 186--225, nov 2016.
\newblock ISSN 0168-9002.
\newblock \doi{10.1016/J.NIMA.2016.06.125}.
\newblock URL
  \url{https://www.sciencedirect.com/science/article/pii/S0168900216306957}.

\bibitem[et~al.(2014)]{Mendoza2014}
E.~Mendoza et~al.
\newblock {New standard evaluated neutron cross section libraries for the
  GEANT4 code and first verification}.
\newblock \emph{IEEE Transactions on Nuclear Science}, 61:\penalty0 2357--2364,
  2014.
\newblock ISSN 00189499.
\newblock \doi{10.1109/TNS.2014.2335538}.

\bibitem[Mendoza et~al.(2018)Mendoza, Cano-Ott, and Capote]{Mendoza2018}
E.~Mendoza, D.~Cano-Ott, and R.~Capote.
\newblock { Update of the Evaluated Neutron Cross Section Libraries for the
  Geant4 Code, IAEA technical report INDC(NDS)-0758 (releases JEFF-3.3,
  JEFF-3.2, ENDF/B-VIII.0, ENDF/B-VII.1, BROND-3.1 and JENDL-4.0u)}.
\newblock Technical report, CIEMAT, Vienna, 2018.
\newblock URL
  \url{https://www-nds.iaea.org/geant4/figures/G4\_10.04.p01\_VS\_MCNP6\_ENDF80.pdf}.

\bibitem[et~al.(2018{\natexlab{a}})]{Tran2018a}
H.~N.~Tran et~al.
\newblock {Comparison of the thermal neutron scattering treatment in MCNP6 and
  GEANT4 codes}.
\newblock \emph{Nuclear Instruments and Methods in Physics Research Section A:
  Accelerators, Spectrometers, Detectors and Associated Equipment},
  893:\penalty0 84--94, jun 2018{\natexlab{a}}.
\newblock ISSN 01689002.
\newblock \doi{10.1016/j.nima.2018.02.094}.
\newblock URL
  \url{https://linkinghub.elsevier.com/retrieve/pii/S0168900218302651}.

\bibitem[Thulliez et~al.(2022)Thulliez, Jouanne, and Dumonteil]{Thulliez2022}
L.~Thulliez, C.~Jouanne, and E.~Dumonteil.
\newblock {Improvement of Geant4 Neutron-HP package: From methodology to
  evaluated nuclear data library}.
\newblock \emph{Nuclear Instruments and Methods in Physics Research Section A:
  Accelerators, Spectrometers, Detectors and Associated Equipment},
  1027:\penalty0 166187, 3 2022.
\newblock ISSN 01689002.
\newblock \doi{10.1016/j.nima.2021.166187}.

\bibitem[et~al.(2015)]{Brun2015}
E.~Brun et~al.
\newblock {TRIPOLI-4®, CEA, EDF and AREVA reference Monte Carlo code}.
\newblock \emph{Annals of Nuclear Energy}, 82:\penalty0 151--160, 8 2015.
\newblock ISSN 03064549.
\newblock \doi{10.1016/j.anucene.2014.07.053}.

\bibitem[et~al.(2018{\natexlab{b}})]{Werner2018}
C.~J.~Werner et~al.
\newblock {MCNP Version 6.2 Release Notes}, 2 2018{\natexlab{b}}.

\bibitem[Cullen(1979)]{Cullen1979}
D.~E. Cullen.
\newblock {Program SIGMA1 (version 79-1): Doppler broaden evaluated cross
  sections in the evaluated nuclear data file/version B (ENDF/B) format}, 10
  1979.

\bibitem[et~al.(2019)]{intercomp}
I.~Duhamel et~al.
\newblock {International Criticality Benchmark Comparison for Nuclear Data
  Validation}.
\newblock \emph{Transactions of the American Nuclear Society}, 121, 2019.

\bibitem[et~al.(2011)]{Chadwick2011}
M.~B.~Chadwick et~al.
\newblock {ENDF/B-VII.1 Nuclear Data for Science and Technology: Cross
  Sections, Covariances, Fission Product Yields and Decay Data}.
\newblock \emph{Nuclear Data Sheets}, 112:\penalty0 2887--2996, 12 2011.
\newblock ISSN 00903752.
\newblock \doi{10.1016/j.nds.2011.11.002}.

\bibitem[Zmeškal and Thulliez(2023)]{MarekGitLab2023}
M.~Zmeškal and L.~Thulliez.
\newblock {Geant4 Neutron-HP benchmarking application}, 2023.
\newblock URL
  \url{https://gitlab.com/lthullie/geant4\_neutronhp\_benchmarkapp}.

\bibitem[Coveyou et~al.(1956)Coveyou, Bate, and Osborn]{Coveyou1956}
R.~R. Coveyou, R.~R. Bate, and R.~K. Osborn.
\newblock {Effect of moderator temperature upon neutron flux in infinite,
  capturing medium}.
\newblock \emph{Journal of Nuclear Energy (1954)}, 2:\penalty0 153--167, 1
  1956.
\newblock ISSN 08913919.
\newblock \doi{10.1016/0891-3919(55)90030-9}.

\bibitem[Becker et~al.(2009)Becker, Dagan, and Lohnert]{Becker2009}
B.~Becker, R.~Dagan, and G.~Lohnert.
\newblock {Proof and implementation of the stochastic formula for ideal gas,
  energy dependent scattering kernel}.
\newblock \emph{Annals of Nuclear Energy}, 36:\penalty0 470--474, 5 2009.
\newblock ISSN 03064549.
\newblock \doi{10.1016/j.anucene.2008.12.001}.

\bibitem[Dagan and Broeders(2006)]{Dagan2006}
R.~Dagan and C.~H.~M. Broeders.
\newblock {On the effect of Resonance dependent Scattering-kernel on Fuel cycle
  and inventory}.
\newblock 9 2006.

\bibitem[Dagan et~al.(2011)Dagan, Becker, Danon, Rapp, Barry, and
  Lohnert]{Dagan2011}
R.~Dagan, B.~Becker, Y.~Danon, M.~Rapp, D.~Barry, and G.~Lohnert.
\newblock {Modelling Resonance Dependent Angular Distribution via DBRC in Monte
  Carlo Codes}.
\newblock \emph{Journal of the Korean Physical Society}, 59:\penalty0 983--986,
  8 2011.
\newblock ISSN 0374-4884.
\newblock \doi{10.3938/jkps.59.983}.

\bibitem[Gunsing and et~al.(2012)]{Gunsing2012}
F.~Gunsing and E.~Berthoumieux et~al.
\newblock {Measurement of resolved resonances of $^{232}$Th(n,$\gamma$) at the
  n\_TOF facility at CERN}.
\newblock \emph{Physical Review C}, 85, 6 2012.
\newblock ISSN 1089490X.
\newblock \doi{10.1103/PHYSREVC.85.064601}.

\bibitem[Rothenstein and Dagan(1998)]{Rothenstein1998}
W.~Rothenstein and R.~Dagan.
\newblock Ideal gas scattering kernel for energy dependent cross-sections.
\newblock \emph{Annals of Nuclear Energy}, 25:\penalty0 209--222, 3 1998.
\newblock ISSN 03064549.
\newblock \doi{10.1016/S0306-4549(97)00063-7}.
\newblock URL
  \url{https://linkinghub.elsevier.com/retrieve/pii/S0306454997000637}.

\bibitem[Rothenstein(2004)]{Rothenstein2004}
W.~Rothenstein.
\newblock Proof of the formula for the ideal gas scattering kernel for nuclides
  with strongly energy dependent scattering cross sections.
\newblock \emph{Annals of Nuclear Energy}, 31:\penalty0 9--23, 1 2004.
\newblock ISSN 03064549.
\newblock \doi{10.1016/S0306-4549(03)00216-0}.
\newblock URL
  \url{https://linkinghub.elsevier.com/retrieve/pii/S0306454903002160}.

\bibitem[Rothenstein(1994)]{Rothenstein1994a}
W.~Rothenstein.
\newblock {Proceedings of the ENS Conference Tel Aviv}, 1994.

\bibitem[Rothenstein and Dagan(1995)]{Rothenstein1995}
W.~Rothenstein and R.~Dagan.
\newblock {Two-body kinetics treatment for neutron scattering from a heavy
  Maxwellian gas}.
\newblock \emph{Annals of Nuclear Energy}, 22:\penalty0 723--730, 11 1995.
\newblock ISSN 03064549.
\newblock \doi{10.1016/0306-4549(95)00002-A}.

\bibitem[Sunny et~al.(2012)Sunny, Brown, Kiedrowski, and Martin]{Sunny2012}
E.~E. Sunny, F.~B. Brown, B.~C. Kiedrowski, and W.~Martin.
\newblock {Temperature effects of resonance scattering for epithermal neutrons
  in MCNP}.
\newblock pages 803--811, 4 2012.

\bibitem[Zoia et~al.(2013)Zoia, Brun, Jouanne, and Malvagi]{Zoia2013}
A.~Zoia, E.~Brun, C.~Jouanne, and F.~Malvagi.
\newblock {Doppler broadening of neutron elastic scattering kernel in
  Tripoli-4®}.
\newblock \emph{Annals of Nuclear Energy}, 54:\penalty0 218--226, 4 2013.
\newblock ISSN 03064549.
\newblock \doi{10.1016/j.anucene.2012.11.023}.

\bibitem[Brown(2019)]{Brown2019}
F.~B. Brown.
\newblock {Doppler Broadening Resonance Correction for Free-gas Scattering in
  MCNP6.2}, 5 2019.

\bibitem[Dagan(2005)]{Dagan2005}
R.~Dagan.
\newblock {On the use of S($\alpha$, $\beta$) tables for nuclides with well
  pronounced resonances}.
\newblock \emph{Annals of Nuclear Energy}, 32:\penalty0 367--377, 3 2005.
\newblock ISSN 03064549.
\newblock \doi{10.1016/j.anucene.2004.11.003}.

\bibitem[Lee et~al.(2009)Lee, Smith, and Rhodes]{Lee2009}
D.~Lee, K.~Smith, and J.~Rhodes.
\newblock {The impact of 238U resonance elastic scattering approximations on
  thermal reactor Doppler reactivity}.
\newblock \emph{Annals of Nuclear Energy}, 36:\penalty0 274--280, 4 2009.
\newblock ISSN 03064549.
\newblock \doi{10.1016/j.anucene.2008.11.026}.

\bibitem[Ouisloumen and Sanchez(1991)]{Ouisloumen1991}
M.~Ouisloumen and R.~Sanchez.
\newblock {A Model for Neutron Scattering Off Heavy Isotopes That Accounts for
  Thermal Agitation Effects}.
\newblock \emph{Nuclear Science and Engineering}, 107:\penalty0 189--200, 3
  1991.
\newblock ISSN 0029-5639.
\newblock \doi{10.13182/NSE89-186}.

\bibitem[Ghrayeb et~al.(2011)Ghrayeb, Ouisloumen, Ougouag, and
  Ivanov]{Ghrayeb2011}
S.~Z. Ghrayeb, M.~Ouisloumen, A.~M. Ougouag, and K.~N. Ivanov.
\newblock {Deterministic modeling of higher angular moments of resonant neutron
  scattering}.
\newblock \emph{Annals of Nuclear Energy}, 38:\penalty0 2291--2297, 10 2011.
\newblock ISSN 03064549.
\newblock \doi{10.1016/j.anucene.2011.04.025}.

\bibitem[Macfarlane et~al.(2017)Macfarlane, Muir, Boicourt, Kahler, and
  Conlin]{Macfarlane2017}
R.~Macfarlane, D.~W. Muir, R.~M. Boicourt, A.~C. Kahler, and J.~L. Conlin.
\newblock {The NJOY Nuclear Data Processing System, Version 2016}, 1 2017.

\bibitem[Cai et~al.(2014)Cai, Llamas-Jansa, and Hauback]{Cai2014}
X.~X. Cai, I.~Llamas-Jansa, and B.~C. Hauback.
\newblock {Accuracy and speed of the HP models}.
\newblock 7 2014.
\newblock URL
  \url{https://indico.cern.ch/event/319884/contributions/1698321/attachments/616390/848181/Geant4\_neutron\_workshop.pdf}.

\bibitem[Yesilyurt et~al.(2012)Yesilyurt, Martin, and Brown]{Yesilyurt2012}
G.~Yesilyurt, W.~R. Martin, and F.~B. Brown.
\newblock {On-the-Fly Doppler Broadening for Monte Carlo Codes}.
\newblock \emph{Nuclear Science and Engineering}, 171:\penalty0 239--257, 7
  2012.
\newblock ISSN 0029-5639.
\newblock \doi{10.13182/NSE11-67}.

\bibitem[Martin et~al.(2014)Martin, Brown, Wilderman, and
  Yesilyurt]{Martin2014}
W.~R. Martin, F.~B. Brown, S.~Wilderman, and G.~Yesilyurt.
\newblock {On-The-Fly Neutron Doppler Broadening in MCNP}.
\newblock page 03102. EDP Sciences, 6 2014.
\newblock ISBN 978-2-7598-1269-1.
\newblock \doi{10.1051/snamc/201403102}.

\bibitem[Forget et~al.(2014)Forget, Xu, and Smith]{Forget2014}
B.~Forget, S.~Xu, and K.~Smith.
\newblock {Direct Doppler broadening in Monte Carlo simulations using the
  multipole representation}.
\newblock \emph{Annals of Nuclear Energy}, 64:\penalty0 78--85, 2 2014.
\newblock ISSN 03064549.
\newblock \doi{10.1016/j.anucene.2013.09.043}.

\bibitem[Viitanen and Leppänen(2012)]{Viitanen2012}
T.~Viitanen and J.~Leppänen.
\newblock {Explicit Treatment of Thermal Motion in Continuous-Energy Monte
  Carlo Tracking Routines}.
\newblock \emph{Nuclear Science and Engineering}, 171:\penalty0 165--173, 6
  2012.
\newblock ISSN 0029-5639.
\newblock \doi{10.13182/NSE11-36}.

\end{thebibliography}





\end{document}